\documentstyle[12pt,aaspp4]{article}

\def\kms{km s$^{\rm -1}$}

\begin{document}

\title{Rotation and Macroturbulence in Metal-poor Field
Red Giant and Red Horizontal Branch Stars}

\author{Bruce W.\ Carney\altaffilmark{1}}
\affil{Department of Physics \& Astronomy\\University of North Carolina\\
Chapel Hill, NC 27599-3255\\e-mail: bruce@unc.edu\\}
\author{David F.\ Gray}
\affil{Department of Astronomy\\University of Western Ontario\\
London, ON N6A 3K7, Canada\\email:dfgray@uwo.ca\\}
\author{David Yong\altaffilmark{2}}
\affil{Department of Physics \& Astronomy\\University of North Carolina\\
Chapel Hill, NC 27599-3255\\e-mail: yong@physics.unc.edu\\}
\author{David W. Latham}
\affil{Harvard-Smithsonian Center for Astrophysics\\60 Garden Street,
Cambridge, MA  02138\\e-mail: dlatham@cfa.harvard.edu\\}
\author{Nadine Manset \& Rachel Zelman}
\affil{Canada-France-Hawaii Telescope Corporation\\65-1238 Mamalahoa
Highway\\Kamuela, HI 96743\\email: manset@cfht.hawaii.edu; zelman@cfht.hawaii.edu}
\author{John B. Laird}
\affil{Department of Physics \& Astronomy\\Bowling Green State University\\
Bowling Green, OH 43403\\e-mail: laird@bgsu.edu\\}

\altaffiltext{1}{Based on observations obtained at the Canada-France-Hawaii
Telescope (CFHT) which is operated by the National Research Council
of Canada, the Institut National des Sciences de l'Univers of the 
Centre National de la
Recherche Scientifique de France, and the University of Hawii.}

\altaffiltext{2}{Mt.\ Stromlo Observatory\\Cotter Road, Weston Creek\\
Canberra, ACT 72611, Australia\\email: yong@mso.anu.edu.au\\}

\begin{abstract}
We report the results for rotational velocities,
$V_{\rm rot}$~sin~$i$, and macroturbulence dispersions,
$\zeta_{\rm RT}$,
for 12 metal-poor field red giant branch (RGB) stars and 7 metal-poor
field red horizontal branch (RHB) stars. The results
are based on Fourier transform analyses of absorption
line profiles from high-resolution
($R \approx\ 120,000$), high-S/N ($\approx\ 215$ per pixel; 
$\approx\ 345$ per resolution
element) spectra obtained with the Gecko spectrograph at CFHT.
The stars were selected from the studies of 20 RHB and 116 RGB
stars from Carney et al.\ (2003, 2007), based primarily on
larger-than-average line broadening values. We find that
$\zeta_{\rm RT}$ values for the metal-poor RGB stars are very
similar to those for metal-rich disk giants studied earlier by
Gray and his collaborators. Six of the RGB stars have small rotational
values, less than 2.0 \kms, while five show significant 
rotation/enhanced line broadening, over 3 \kms. We confirm the
rapid rotation rate for RHB star HD~195636, found earlier by Preston (1997).
This star's rotation is comparable to that of the fastest known
rotating blue horizontal branch (BHB) stars, when allowance is
made for differences in radii and moments of inertia. The
other six RHB stars have somewhat lower rotation but show
a trend to higher values at higher temperatures
(lower radii). Comparing our results with those for BHB stars
from Kinman et al.\ (2000), we find that the fraction of rapidly rotating
RHB stars is somewhat lower than is found among BHB stars. 
The number of rapidly rotating RHB
stars is also smaller than we would have
expected from the observed rotation of
the RGB stars. We devise two
empirical methods to translate the line broadening results
obtained by Carney et al.\ (2003, 2007) into $V_{\rm rot}$~sin~$i$
for all the RGB and RHB stars they studied. Binning the RGB
stars by luminosity, we find that most metal-poor field RGB
stars show no detectable sign, on average, of rotation, which
is not surprising given the stars' large radii. However,
the most luminous stars, with $M_{\rm V} \leq\ -1.5$,
do show net rotation, with mean values of 2 to 4 \kms, depending
on the algorithm employed, and also show signs of radial
velocity jitter and mass loss.
This ``rotation" may in fact prove
to be due to other line broadening effects, such as
shock waves or pulsation.
\end{abstract}


\section{INTRODUCTION}

In previous papers (Carney et al.\ 2003, Carney et al.\ 2007; 
hereafter C2003, C2007)
we reported on radial velocities and line broadenings for 
136 metal-poor field red
giant branch (RGB) and red horizontal branch (RHB) stars, 
based on 2413 high-resolution, low-S/N
spectra. One of the more intriguing results was that the 
more luminous red giants,
as well as many of their evolutionary progeny, red horizontal branch stars,
showed significant line broadening. Interpreting the enhanced line broadening
as rotation, C2003 explored the possibility that it might 
have arisen from 
absorption of one (or more) jovian mass planets that were engulfed only
as the red giants swelled to large enough radii. 

C2003 suggested several follow-up studies. First, the sample size should be
expanded, and C2007 presented the results for 45 stars to complement the
original 91-star sample of C2003. Second, high-precision radial velocity
monitoring of metal-poor dwarfs and subgiants should be undertaken to
explore the frequency of jovian-mass planets with orbital periods of
order one year, corresponding to aphelion distances comparable to the
maximum radial size of metal-poor RGB stars. The initial results
of such a study have been reported (Sozzetti et al.\ 2006), and it appears
that such planetary companions are not sufficiently common to explain
the modest frequency of significant line broadening among the most luminous
metal-poor red giants. 

This paper explores the separate contributions of rotation and
macroturbulence, based on a selected subsample of stars
studied by C2003 and C2007. For example, if 
macrotubulence is a strong function of luminosity,
the enhanced line broadenings found by C2003 and C2007 among the field
red giants might be explained. If macroturbulence is a strong function
of temperature, perhaps the line broadening seen in some of the red
horizontal branch stars could also be explained. But if rotation
is the cause of the enhanced line broadening among the stars with the
largest radii, some new explanations must be sought.

To test these hypotheses, we decided to exploit methods developed by
Gray (1982), whereby line profiles measured using very 
high-resolution, high-S/N spectra could be analyzed via Fourier transform
methods to distinguish the contributions of rotation and macroturbulence.

\section{PROGRAM DESCRIPTION}

\begin{figure}
\epsscale{0.80}
\plotone{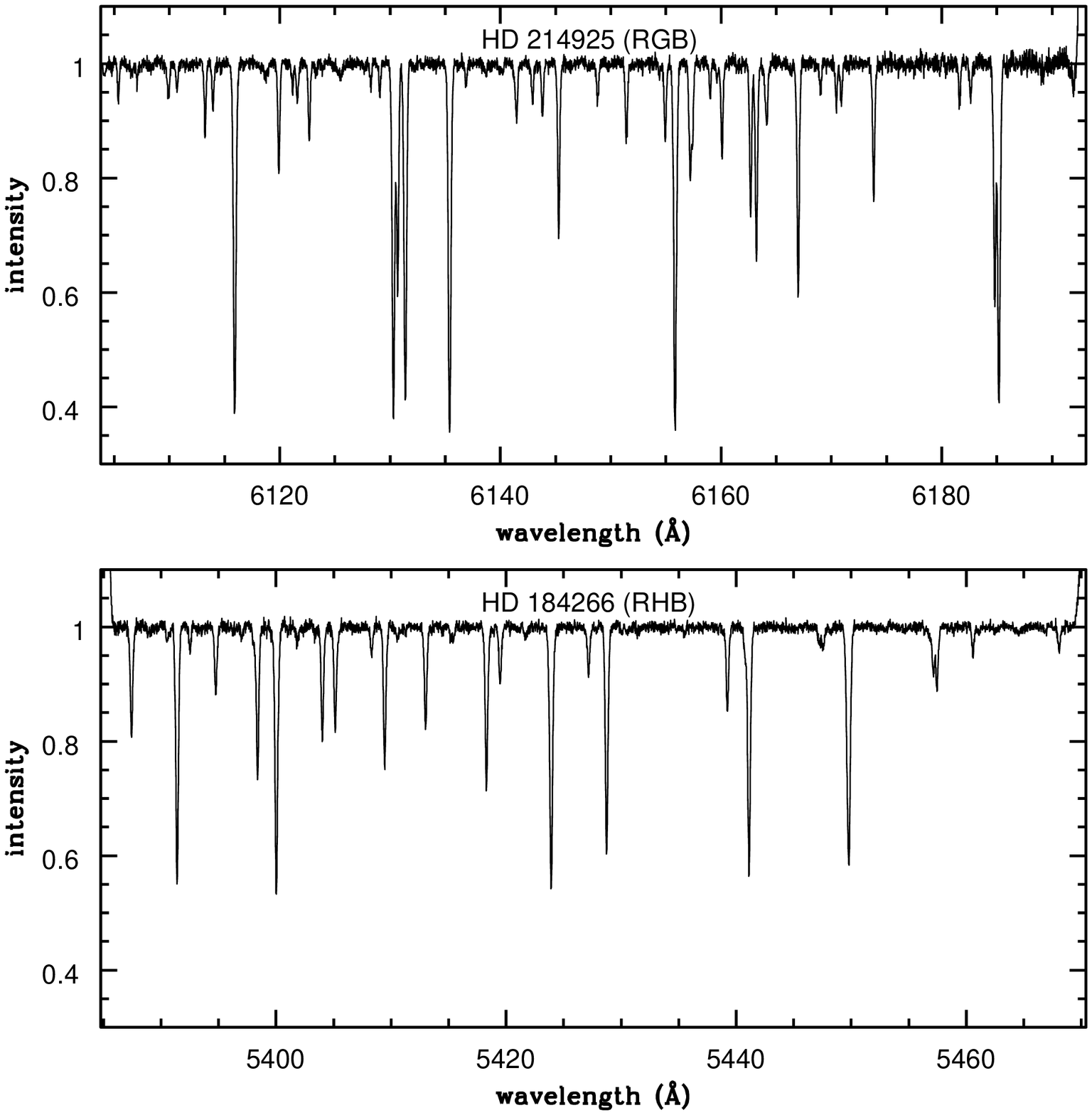}
\caption{The spectral coverage for one of our twelve red giant branch stars
and one of the seven red horizontal branch stars.
\label{fig:spectra}}
\end{figure}

The Fourier transform method requires very high-resolution and high-S/N
spectra. The Gecko spectrograph on the Canada-France-Hawaii Telescope
(CFHT) was deemed to be an ideal instrument for our work, but the
wavelength coverage is quite limited. We therefore computed a grid
of model atmospheres covering the stellar parameters appropriate
to our field RHB and RGB stars, using ATLAS9.
We then computed synthetic spectra using R.\ L.\ Kurucz's code SYNTHE.
We sought wavelength regions that had a significant number of uncrowded
absorption lines. The lines must be reasonably strong, but not saturated
since pressure-broadened line wings render the lines less useful. We
determined that for the red horizontal branch stars, the optimal
wavelength region should be centered at 5430~\AA, while for the
red giants, the central region should be 6150~\AA. Figure~\ref{fig:spectra}
shows the spectra for one of the RGB and one of the RHB stars.

Because of the requirement for high-S/N, and limited available
observing time, we had to choose our targets carefully. Of course
we selected a number of RHB and RGB stars with significant line
broadening. We also elected to observe a few stars with
smaller line broadening, partly as a test of the line broadening
derived from the lower-resolution ($R \approx\ 32,000$)
CfA spectra. Further, if the line broadening is due to rotation, we 
assume that the less-broadened stellar spectra might reflect
nearly pole-on inclinations, 
so that we could explore
macroturbulence more carefully. 

Our observations were obtained in two runs with the CFHT, and
to check the consistency of our results,
we observed HD~29574 during both. We also
felt a need to compare our
results with the extensive studies of disk stars completed earlier by
Gray (1982, 1984), Gray \& Toner
(1986, 1987), and Gray \& Pallavicini (1989). 
We therefore included $\eta$~Ser (HR~6869) in our program, which
had been studied previously by Gray \& Pallavicini (1989).

\section{OBSERVATIONS}

We used the Gecko spectrograph at the CFHT on two observing runs,
in December 2004 and October 2006. We used the MIT2 detector,
a thinned 2048 $\times$ 4096 chip with 15$\mu$ pixels. The
read noise for this device is about 7.5 electrons, which was
negligible given the strong exposures. The gain setting
was 1.2 electrons per ADU. Gecko was fiber-fed by CAFE (Baudran \&
Vitry 2000) from the Cassegrain focus of the telescope. Fiber modal
noise was suppressed by agitating the fiber continuously 
(Baudrand and Walker 2001).

\begin{figure}
\epsscale{0.80}
\plotone{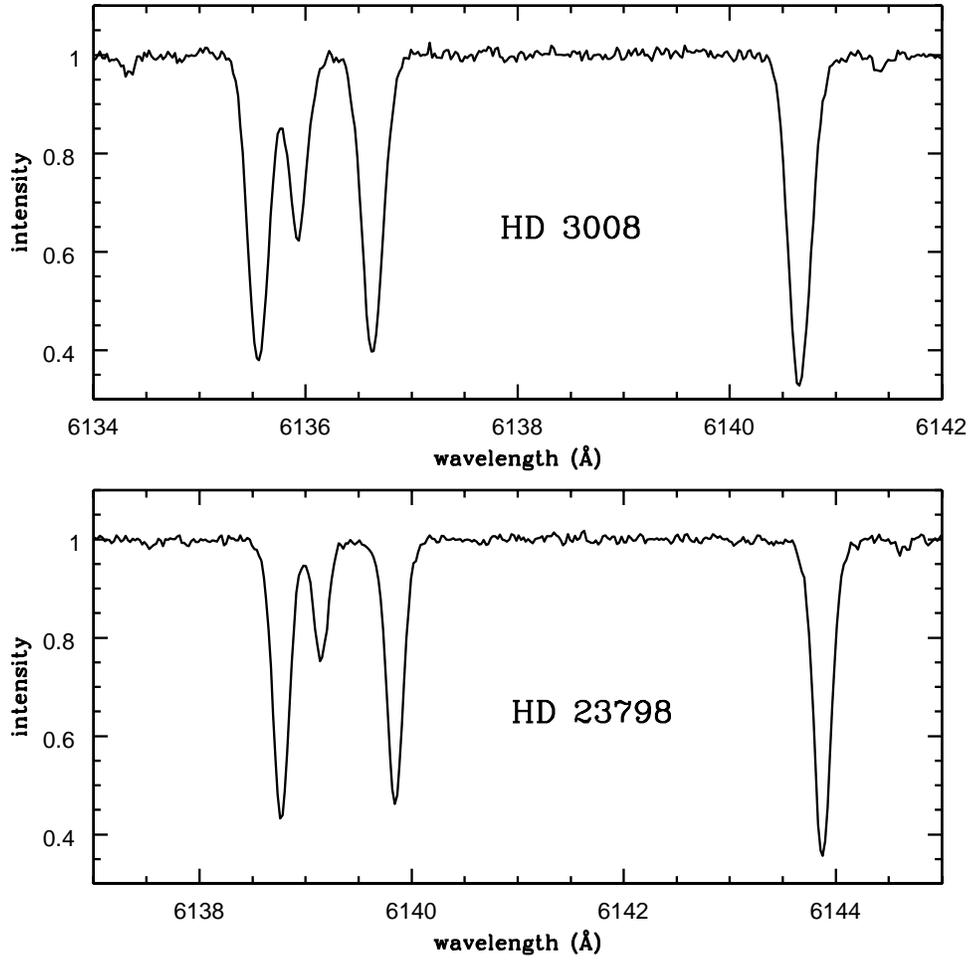}
\caption{A ``close-up" of the CFHT spectra of two of our program
red giant branch stars. HD~3008 has slightly broader lines, as
the analyses of the CfA and CFHT spectra indicated.
\label{fig:rgb}}
\end{figure}

The RGB stars were observed using order 9 and the 1521 filter.
The single order on the detector at 6150 \AA\
spans only about 90 \AA, and the dispersion is about 1.47 \AA\ mm$^{-1}$
(0.022 \AA\ pixel$^{-1}$). The RHB stars were observed using
order 10 and the 1510 filter, which 
covered 86 \AA\ at a dispersion of 1.40 \AA\ mm$^{-1}$, or about
0.021 \AA\ pixel$^{-1}$. 
We measured the resolution using Th-Ar
comparison lines, finding a typical resolving power of 120,000.
Figure~\ref{fig:rgb} shows 8 \AA\ coverage in spectra of
two red giant branch stars. The line depths are comparable,
and it is (marginally) apparent that HD~3008 is broader lined
than HD~23798, as the analyses of both the CfA and
the CFHT spectra revealed.

Table~1 provides a log of our observations, including 
the exposure time in minutes, the heliocentric Julian date
of mid-exposure, and the
estimated signal-to-noise obtained {\em per pixel}. Each
spectral resolution element covered about 2.3 pixels, so the
S/N per resolution element is that given in Table~1 multiplied
by a factor of about 1.5.

\section{DERIVATION OF ROTATIONAL AND MACROTURBULENT VELOCITIES
FOR CFHT PROGRAM STARS}

Since the Doppler broadening of rotation and 
macroturbulence are comparable in size, 
it is necessary to push toward high Fourier frequencies 
in order to distinguish the 
subtle differences in shape they impress upon 
the spectral lines.  This is why the 
high resolving power of the Gecko spectrograph 
was needed.  But high resolving power 
alone is not sufficient because the amplitudes 
at high Fourier frequencies are 
small and often below the noise level.  
For this reason high signal-to-noise ratios 
are also needed.  Most of our observations are 
of sufficient quality to fulfill these 
requirements and allow us to distinguish rotation from macroturbulence.

Individual line profiles were extracted and 
corrected for small blends when necessary.  
The Fourier analysis then proceeds in the usual 
way (Gray 2005) 
by first dividing out the transform of a thermal profile computed 
from a model photosphere.  
Effective temperatures, surface gravities, and 
metallicities were taken from C2003 and C2007.
Treatment of the thermal profile is not overly critical since 
its width is considerably 
smaller than the observed line widths.  When this 
step is completed for all the usable 
lines, an average of these ratios is taken.  
The final manipulation of the data is to 
divide out the transform of the instrumental profile.  
We took the profile of a narrow 
emission line in a thorium-argon comparison 
lamp to be the instrumental profile.  
While we would have preferred using a narrower-line 
source, none was available.  
However, we expect no serious error to be introduced 
because the instrumental profile 
is many times narrower than the stellar lines.  
Since both the transforms of 
the thermal profile and the instrumental 
profile decline toward larger frequencies, division by 
them enhances the noise at
the high frequencies.  
The transition to enhanced noise is 
fairly abrupt.  Naturally, our analysis is restricted
to Fourier frequencies below this transition.

\begin{figure}
\plotone{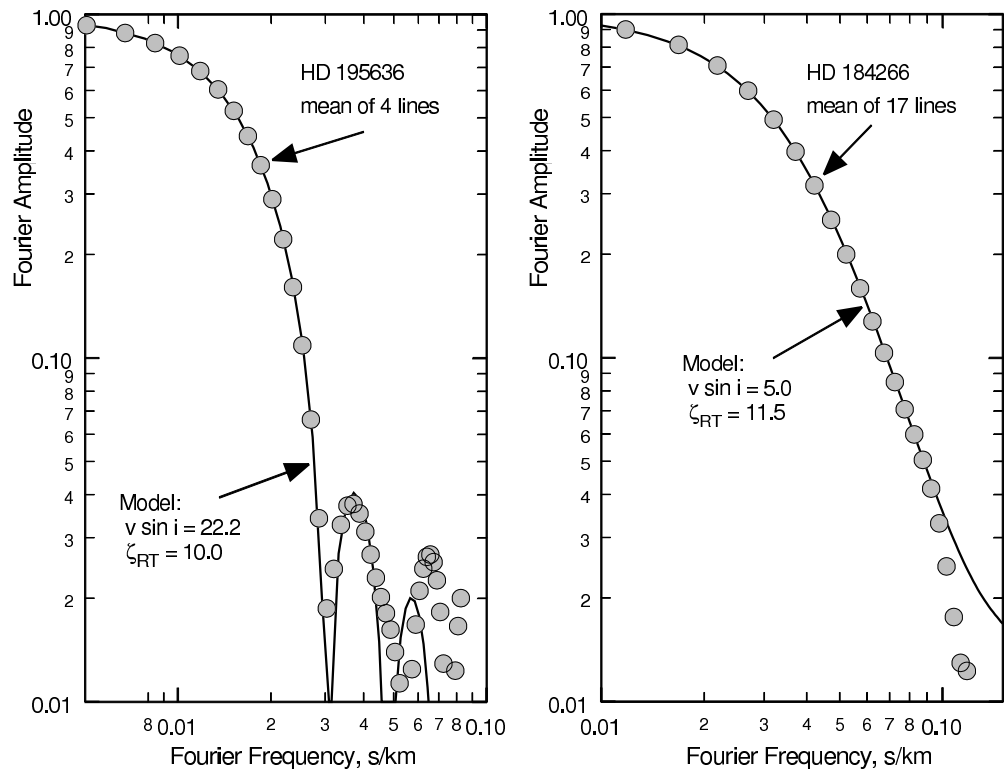}
\caption{The mean residual
transforms (circles) are shown with the adopted models (line) for two
of our program stars.
\label{fig:transfig}}
\end{figure}

The distribution of Doppler 
shifts from rotation and Radial-Tangential macroturbulence 
(Gray 2005) are computed by integrating 
over a model stellar disk on a sector-annulus 
format.  A sector step of 0.5 degree 
was used and the annulus dimension was adjusted to 
be of comparable linear dimension.  
A limb darkening coefficient ($\epsilon$ = 0.7) was used.  
Fourier transforms of these Doppler-shift 
distributions are compared with the observations, and
the broadening parameters, $V_{\rm rot}$~sin~$i$ and 
$\zeta_{\rm RT}$\footnote{$\zeta_{\rm RT}$ is the radial-tangential
macroturbulence dispersion. It is not the Gaussian
macroturbulent velocity, but is roughly 2.4 times
larger (Gray 1978).}, are adjusted until the best match is
obtained.
The ratio of rotational to macroturbulence broadening, 
$V_{\rm rot}$~sin~$i$/$\zeta_{\rm RT}$, is determined 
from the curvature and any sidelobe structure.  
The absolute scale of the velocities 
comes from the translational match on the 
logarithmic abscissa.  In those cases where 
$V_{\rm rot}$~sin~$i$ is considerably smaller 
than $\zeta_{\rm RT}$, $V_{\rm rot}$~sin~$i$ 
will be poorly determined, and 
vice versa.  We estimated the errors by 
altering the $V_{\rm rot}$~sin~$i$ and $\zeta_{\rm RT}$ parameters by 
small amounts until obvious mismatch with the data occurs.
Two examples of the final step are shown in Figure~\ref{fig:transfig}.
HD~195636 is a rapid rotator, while HD~184266 has some
rotation but larger macroturbulence.
Table~\ref{tab2} summarizes the results of the Fourier analyses.

Our results appear to be consistent between the two observing
runs, based on the very good agreement for the two sets of
measurements of HD~29574. Further, our results for $\eta$~Ser,
$V_{\rm rot}$~sin~$i$ = $1.0 \pm 0.8$ \kms and
$\zeta_{\rm RT}$ = $4.1 \pm 0.5$ \kms, agree very well
with those obtained by Gray \& Pallavicini (1989),
$2.0 \pm 0.5$ and $4.0 \pm 0.5$ \kms, respectively.

\section{RESULTS}

\subsection{Radial Velocities}

The radial velocities for each star reported in Table~1 were
derived using {\bf rvsao} (Kurtz \& Mink 1998) running inside the
IRAF\footnotemark \footnotetext{IRAF (Image Reduction and Analysis
Facility) is distributed by the National Optical Astronomy
Observatories, which are operated by the Association of Universities
for Research in Astronomy, Inc., under contract with the National
Science Foundation.} environment. We compare our results
with those reported in C2003 and C2007 (which therefore excludes
$\eta$~Ser). We do not include HD~218732 in the comparisons because
it is a spectroscopic binary. 

For the 11 stars not known to suffer velocity ``jitter" (see C2003
and C2007 for a more complete discussion of this phenomenon),
we find $<V_{\rm rad,CFHT} - V_{\rm rad,CfA}>$ = $+0.13 \pm 0.14$
\kms, with $\sigma$ = 0.46 \kms. This agreement is very satisfactory.
For the 8 other stars known to be subject to ``jitter", the mean
difference is $-0.71 \pm 0.44$ \kms, with $\sigma$ = 1.24 \kms.
Considering the velocity variations in these stars, this agreement
is good.

\subsection{A First Look at Rotational and Macroturbulent Velocities}

\begin{figure}
\epsscale{0.80}
\plotone{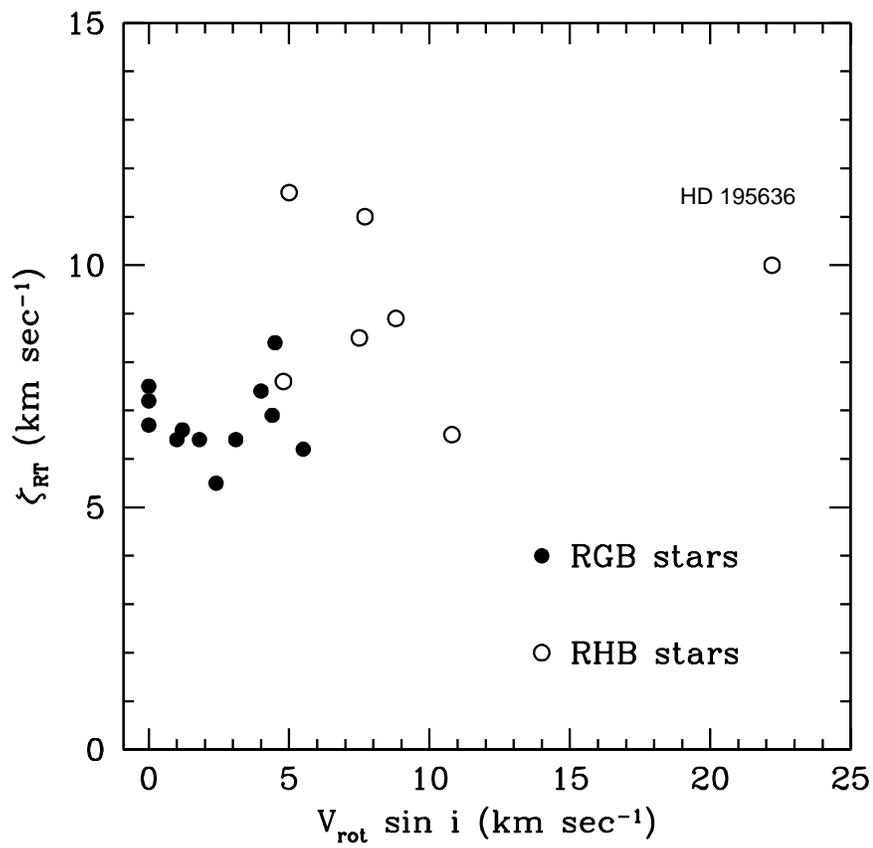}
\caption{A comparison of our derived macroturbulence dispersion
values, $\zeta_{\rm RT}$, and rotational velocities, $V_{\rm rot}$~sin~$i$.
\label{fig:comparezetavrot}}
\end{figure}

Figure~\ref{fig:comparezetavrot} distinguishes the RGB (filled circles)
from the RHB stars (open circles) in the $V_{\rm rot}$~sin~$i$ vs.\
$\zeta_{\rm RT}$ plane. 
Several points are apparent from the Figure.

First, much of the line broadening 
found by C2003 and C2007 for the most
luminous red giants, with 
$V_{\rm broad}$\footnotemark \footnotetext{Note
that C2003 refer to rotational velocities, $V_{\rm rot}$~sin~$i$, but
which we prefer to call broadening velocities, $V_{\rm broad}$.} 
values approaching 12 \kms, is due more to
macroturbulence than to rotation, whose maximal value among the
12 RGB stars we have studied is only 5.5 \kms. The 12 RGB stars
have $<V_{\rm broad,CfA}>$ = $8.1 \pm 0.6$ \kms ($\sigma$ = 2.2 \kms),
but $<V_{\rm rot}$~sin~$i>$ = $2.3 \pm 0.6$ \kms ($\sigma$ = 1.9 \kms),
and $<\zeta_{\rm RT}>$ = $6.8 \pm 0.2$ \kms ($\sigma$ = 0.7 \kms).

Second, the macroturbulence levels are generally higher in the observed RHB stars 
than in the RGB stars, with $<\zeta_{\rm RT}>$ = $9.1 \pm 0.7$ 
($\sigma$ = 1.8 \kms).
The RHB stars have higher gravities than the
RGB stars, so we would be tempted to assume that macroturbulence
increases at higher gravities, smaller radii, or lower luminosities,
but the discussion in Section~6.2.2 reveals the opposite to be the
case. This also conflicts with the findings of Gray (1982) and
Gray \& Pallavicini (1989), who found that lower gravity disk
giants have higher values of $\zeta_{\rm RT}$.
Therefore, temperature must play a significent role as well,
as had been demonstrated earlier by Gray (1982) and by Gray \& Toner (1986).

Third, rotation and macroturbulence play comparable roles in the
line broadening of the observed RHB stars. Including the 
rapid rotator HD~195636, 
$<V_{\rm rot}$~sin~$i>$ = $9.5 \pm 2.2$ \kms\ ($\sigma$ = 6.0 \kms).
Excluding that star, 
$<V_{\rm rot}$~sin~$i>$ = $7.4 \pm 0.9$ \kms\ ($\sigma$ = 2.3 \kms).
These values are comparable to $<\zeta_{\rm RT}>$ for the observed RHB stars.

Fourth, the RHB stars show higher rotational velocities than the RGB
stars, especially in the case of HD~195636, a star whose rapid rotation
was noted first by Preston (1997).
Since RHB stars represent
some of the descendents of RGB stars, and since
RHB stars have smaller radii, more rapid rotation is expected. 
But there is a discrepancy when the results are examined more closely.
Taking simple means, we find that the 12 RGB stars have 
$<R>$ = 68~$R_{\odot}$, while the 7 RHB stars have $<R>$ = 7.5~$R_{\odot}$
(7.2~$R_{\odot}$ if we exclude HD~195636). We have been unable
to find published moments of inertia of RGB and RHB stars, so
we make the assumption that since the core masses and total
masses of both classes of star are similar, then if the envelope
density distributions are similar, the total moment of inertia
will scale as the stellar radii. In this case, we expect the
RHB stars to be rotating
about nine times faster than the mean RGB rotation rate. The ratio
is, in fact, 4.1, if we include HD~195636,
and is 3.2 if we exclude it. While our RGB and RHB samples
were selected with a bias in favor of larger line broadening, this does
not alter our conclusion. In subsection 7.2.2.\ (see Table~5),
we found that the mean $V_{\rm broad}$
value for the twenty most luminous red giants is 7.7 \kms, very similar to
the 8.1 \kms\ for the RGB stars observed at CFHT. So the bias does not
strongly affect the mean rotational value of the luminous red giants.
The same is not true for the RHB stars, however. The seven RHB stars
in Table~2 have $<V_{\rm broad}>$ = 13.4 \kms\ (12.0 \kms\ if we
exclude HD~195636), while the thirteen RHB stars not observed
at CFHT have $<V_{\rm broad}>$ = 6.4 \kms.
Correcting for the bias for the
RHB sample will only lower their mean rotational velocities, increasing
the magnitude of the discrepancy between the expected
ratio of rotational velocities. What might cause the discrepancy?
One possibility is the loss of angular momentum, perhaps
by a vigorous stellar wind or pulsation at the most luminous final
stages of RGB evolution. RHB stars have larger envelopes than
BHB stars, so, presumably, RHB stars lost less mass during the
RGB stage. But as we discuss in Section 7.1.3, it is not
clear that the two classes of HB stars have significantly different
net amounts of angular momentum. 
Finally, and this bears on discussions below, the
rotation we have measured in the most luminous red giants may
reflect a combination of rotation and some other effect, such
as pulsation, that may also result in line broadening.

\subsection{The Case of HD~195636}

The rotation of HD~195636 is much higher
than the other RHB stars. It is, however, not out of line with the
maximal rotation seen in blue horizontal branch stars. We consider
the BHB stars studied by Kinman et al.\ (2000), exluding
BD+32~2188
because the radius derived 
from its log~$g$ value, 11.8~$R_{\odot}$, suggests
it is not a BHB star.
Of the remaining 29 stars,
two have $V_{\rm rot}$~sin~$i$ $\approx\ 40$ \kms,
and estimated radii of about 3.3 $R_{\odot}$. At the estimated
radius of HD~195636, $\approx\ 9.1$ $R_{\odot}$, this would
correspond to a rotational velocity of about 15 \kms, somewhat
smaller than our derived value of 22 \kms. If these largest
rotational velocities simply reflect stars with the most favorable
viewing angles (sin~$i \approx\ 1$), this small sample suggests
that there is no great difference between the maximum values
of rotation in BHB and
RHB stars.

In drawing attention to this star, Preston (1997) explored whether a
close binary companion could have interacted tidally, producing such
a high rotational velocity. Preston's observations 
had limited time coverage, but
displayed no sign of radial velocity variability. C2003 reported
43 radial velocities covering 5086 days (13.9 years), and did not
detect any radial velocity variability. Our additional radial velocity
measure (Table~1) extends the time coverage to 8460 days (23.2 years),
and the star has maintained the same radial velocity. We conclude that
tidal locking in a binary system is not the source of the rapid rotation.
Indeed, of the 20 RHB stars studied by C2003 and C2007,
only one has proven to be a spectroscopic binary (HD~108317), and it
has one of the smallest line broadening measures (C2003), with
$V_{\rm broad,CfA}$ = 5.1 \kms. Tidal locking does not explain the
relatively high rotational velocities seen in some of our RHB stars.

\section{ESTIMATION OF ROTATIONAL VELOCITIES \& MACROTURBULENT DISPERSIONS
FOR A LARGER SAMPLE OF STARS}

We have obtained rotational velocities and macroturbulent dispersions
for 7 of the 20 RHB stars studied by C2003 and C2007, and 12 of the
116 RGB stars. Further, those 12 RGB stars are all near the tip of the
red giant branch. We believe we can extract rotational velocity 
estimates of the other 13 RHB and 104 RGB stars, at least in a
statistical sense. 

We use two basic approaches. In one case, we seek to identify
a means whereby we can reliably estimate $\zeta_{\rm RT}$ as a
function of some parameter, such as absolute magnitude, gravity,
or temperature, and then employ some algorithm 
to remove that contribution to the total
line broadening determined using the CfA spectra. In the second
case, we simply compare the $V_{\rm broad}$ values determined
from the CfA spectra with the rotational velocities obtained from
the CFHT data. This essentially assumes that macroturbulence
is either constant among our program stars or small in
comparison with rotation.

\begin{figure}
\epsscale{0.80}
\plotone{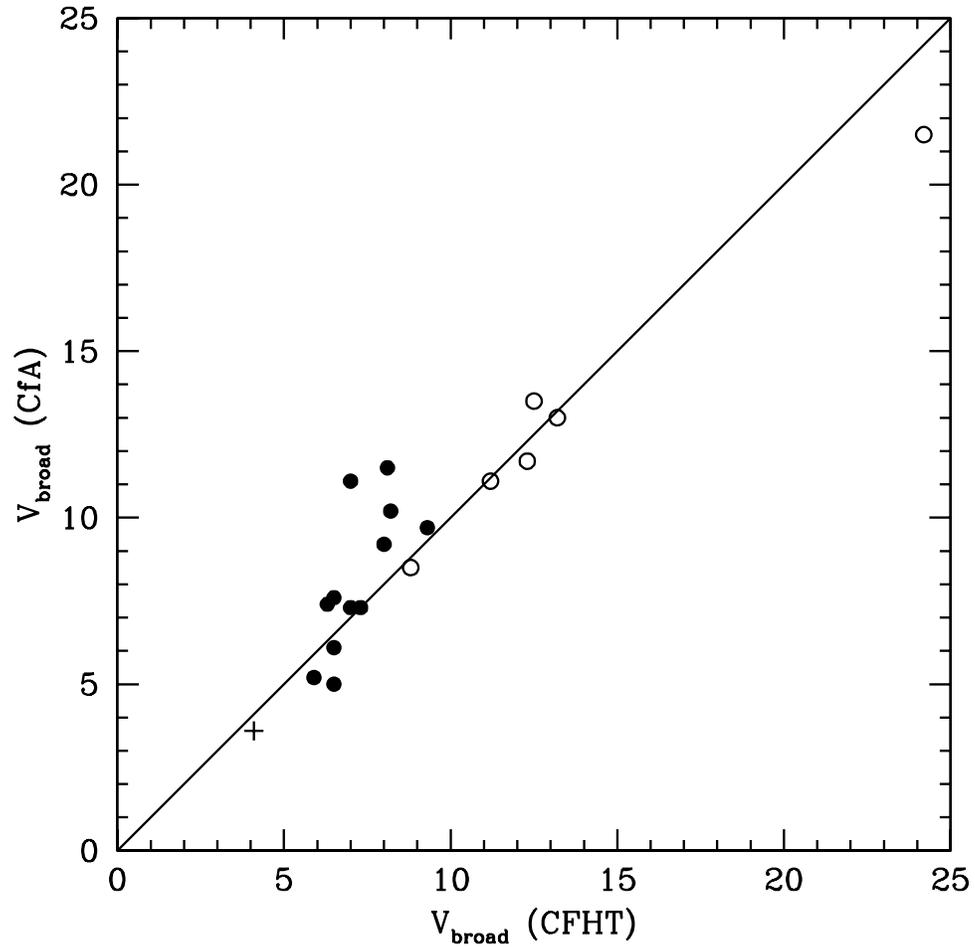}
\caption{A comparison of the line broadening measured using the
CfA spectra, $V_{\rm broad}$ (CfA) with that derived from
our CFHT results using Equation~\ref{eq:vbroad},
$V_{\rm broad}$ (CFHT). Filled circles are RGB stars; open
circles are RHB stars; the plus sign is $\eta$ Ser.
\label{fig:equation1}}
\end{figure}

In the first case, following Massarotti et al.\ (2007), C2007 found that
\begin{equation}
\label{eq:vbroad}
V_{\rm broad}=[(V_{\rm rot} \sin\ i)^{2} + 0.95 \zeta_{\rm RT}^{2}]^{1/2}.
\end{equation}
Figure~\ref{fig:equation1} compares the results from this Equation
using our CFHT data with those derived at CfA.
It is important to recall that this is a purely empirical fit. It is not
based on any deep physical understanding of the phenomena of rotation
and macroturbulence. Indeed, it is hard to justify because $\zeta_{\rm RT}$
and does not behave
as a Gaussian in broadening a stellar absorption line (Gray 2005). We employ the relation
solely on the basis that it appears to provide a good fit to our data.
The 0.95 coefficient was determined empirically by comparing the results
given in Table~2 with the line broadening estimates from the CfA
observations. Specifically, 0.95 was necessary to provide a negligible offset
between $V_{\rm broad,CFHT}$ and $V_{\rm broad,CfA}$. The resultant
scatter in the relation was only 1.1 \kms, and was even smaller, 0.9 \kms,
for stars whose line broadenings were smaller than the resolution of
the CfA spectra (8.5 \kms). 

\begin{figure}
\epsscale{0.80}
\plotone{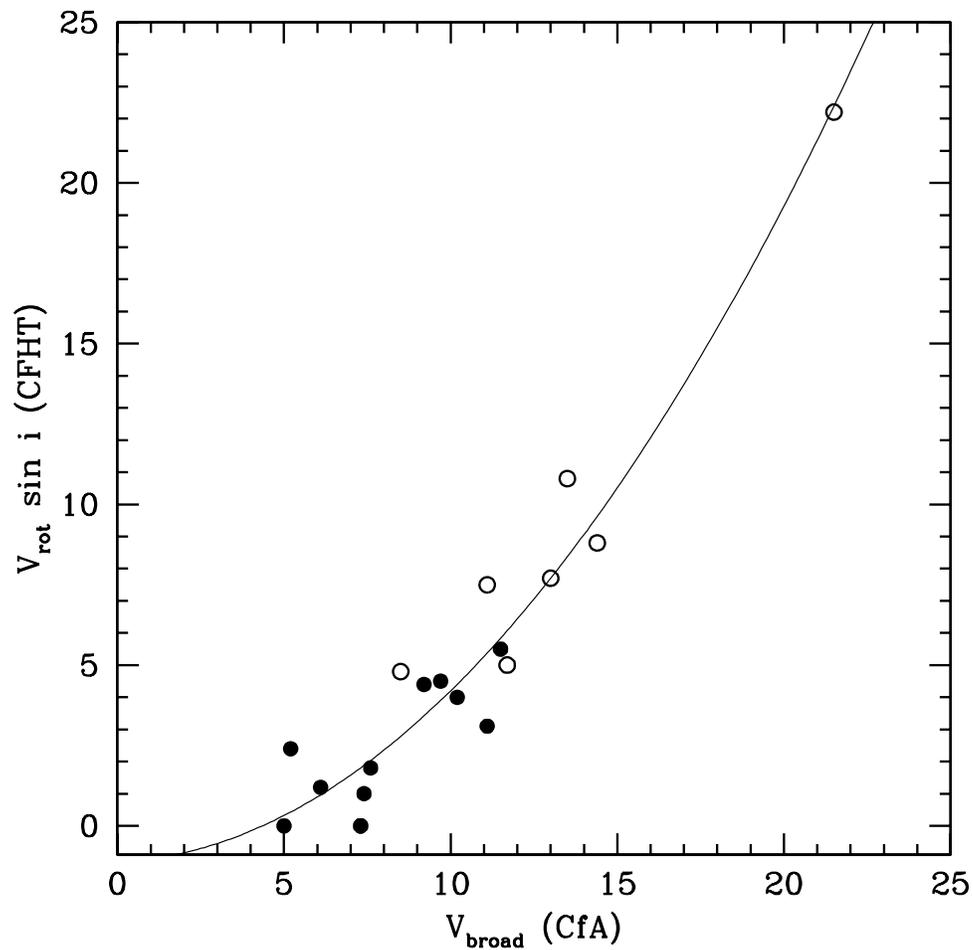}
\caption{A comparison of the rotational velocity, $V_{\rm rot}$~sin~$i$,
derived from the
CfA spectra, $V_{\rm broad}$ (CfA), 
using Equation~\ref{eq:vbroadcorr}, with that measured using the
CFHT spectra, 
$V_{\rm rot}$~sin~$i$ (CFHT). Filled circles are RGB stars; open
circles are RHB stars.
\label{fig:vrotvscfaquadratic} }
\end{figure}

In the second case, we show in 
Figure~\ref{fig:vrotvscfaquadratic} a second-order
fit between the $V_{\rm broad}$ values derived from our CfA spectra
with the $V_{\rm rot}$~sin~$i$ values from the much higher resolution
and much higher S/N CFHT spectra. Filled circles are RGB stars, while
open circles are RHB stars. The fit is remarkably good, with
a scatter of only 1.5 \kms.
\begin{equation}
\label{eq:vbroadcorr}
V_{\rm rot} \sin i = -1.12 + 0.044 V_{\rm broad} + 0.0488 V_{\rm broad}^2
\end{equation}

Note that in both cases we have relied on \underline{both}
the RHB and the RGB stars to calibrate the relations.
We will employ both methods to estimate $V_{\rm rot}$~sin~$i$
for all of our CfA program stars that were not observed at CFHT.

\subsection{Macroturbulence Among the Red Horizontal Branch Stars}

\begin{figure}
\epsscale{0.80}
\plotone{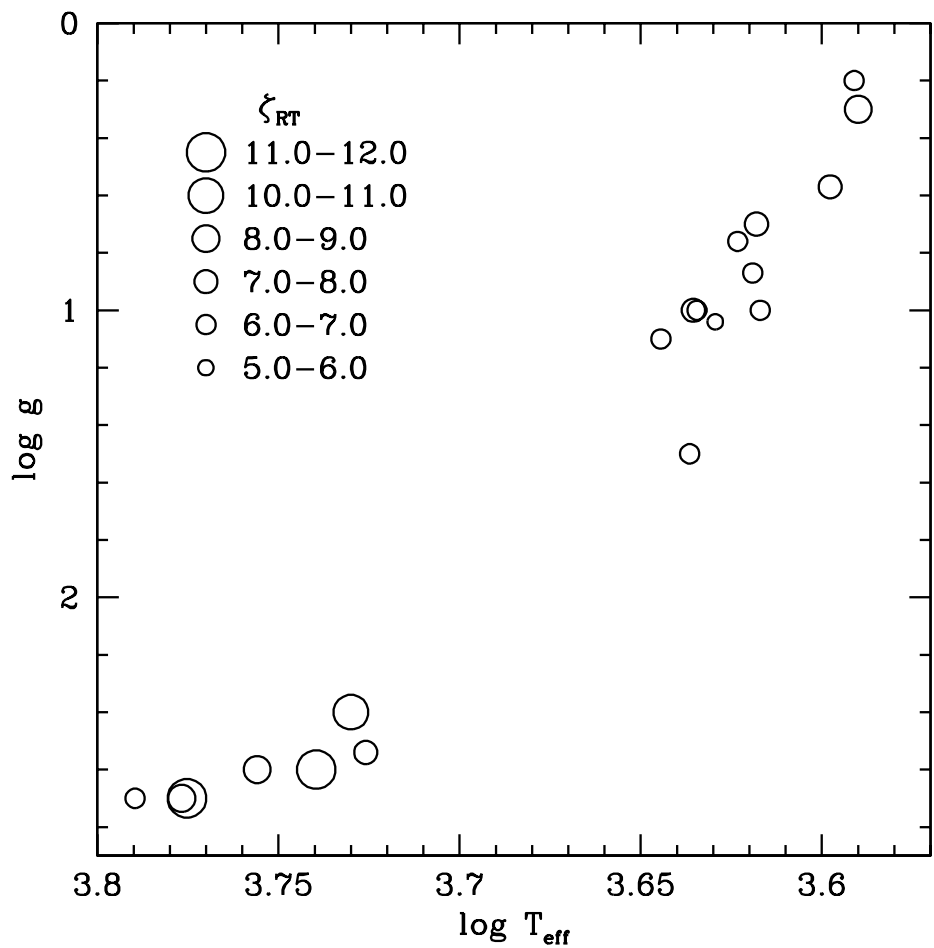}
\caption{A comparison of our derived macroturbulence dispersion
values, $\zeta_{\rm RT}$, as a function of temperature and
gravity. We employ different sizes for the data point to illustrate
the different magnitudes of $\zeta_{\rm RT}$.
\label{fig:zetavstlogg}}
\end{figure}

To make use of Equation~\ref{eq:vbroad}, we need to have good estimates
for the contribution of the macroturbulence to the total line
broadening. Figure~\ref{fig:zetavstlogg} summarizes our macroturbulence
dispersions as a function of temperature and gravity. Let us focus
first on the seven RHB stars at the lower left of the Figure. The
stars have very similar luminosities and gravities, but have
a respectable range in effective temperature, from about 5300~K to
almost 6200~K. However, there is no obvious correlation between
$T_{\rm eff}$ and $\zeta_{\rm RT}$. A weighted least squares fit
results in
\begin{equation}
\label{eq:rhbzeta}
\zeta_{\rm RT} = -0.00158 T_{\rm eff} + 18.17,
\end{equation}
with a correlation coefficient of only $-0.28$. The scatter
about this relation is 1.7 \kms, which is negligibly better
than taking the mean value for all 7 stars, which, as noted
above, results in $<\zeta_{\rm RT}>$ = $9.1 \pm 0.7$ \kms,
with $\sigma$ = 1.8 \kms. In Section~7.1, we return to this
topic, finding an additional reason to doubt that a simple
description of the behavior of $\zeta_{\rm RT}$ can be
applied to RHB stars. On the other hand, the RGB stars
appear to be well-behaved, as we discuss below.

\subsection{Expanding the Sample of Calibrating Red Giants}

As we have pointed out, the red giant stars studied in this
program tend to lie near the tip of the RGB, as may be seen
in Figure~\ref{fig:zetavstlogg}. But the other 104 RGB stars
in C2003 and C2007 have a much wider range of luminosities,
temperatures, and gravities, and hence we require additional
data. Since our results appear to be in good agreement with
those obtained for the disk population giant $\eta$~Ser, we
explore now how the results for our metal-poor field red giants
compare with results obtained earlier for metal-rich 
disk population field red giants.

\subsubsection{Selection of Disk Stars}

The behavior of macroturbulence and rotation of very old and
very metal-poor halo stars merits comparisons with the values
of younger and more metal-rich disk stars. We have assembled
a sample of disk stars from the work of Gray (1982, 1984, 1989), Gray \& Toner
(1986, 1987), and Gray \& Pallavicini (1989), who
employed the same tools to determine $V_{\rm rot}$~sin~$i$ and
$\zeta_{\rm RT}$ from very high-resolution, very high S/N spectra,
as employed here. 
Stars retained in the final comparisons with
the halo giants had to satisfy several criteria.

First, we chose to exclude stars with spectral types earlier
than G8. Nothing is lost thereby since our halo stars have
temperatures cooler than this. The additional benefit is that
the stars we employ are cooler than the transition in rotational 
velocities seen in giant
stars with G spectral types (Gray 1989). Stars cooler than G0 to G3 III
stars have significantly lower rotational velocities
than warmer stars, presumably due to a rotational dynamo-generated
magnetic brake.

We elected to retain only those stars with a
consistently-applied photometric or high-resolution spectroscopic
approach to determinations of stellar parameters. We are fortunate that McWilliam (1990)
undertook an extensive spectroscopic survey of metallicities
of disk giants, and employed consistent photometric estimations
of temperatures and gravities. Fortunately, many of the stars in the studies
by Gray and his colleagues were also studied by McWilliam. We have
adopted his photometric estimates of effective temperature and gravity,
and his spectroscopic determinations of [Fe/H]. However, to avoid
extending the comparisons of disk giants and halo giants to temperatures
far beyond those of our available halo sample giant stars, we have retained only
those stars in McWilliam (1990) with $T_{\rm eff} \leq\ 5000$~K, in order
to be consistent with the use of stars with spectral types of G8 and later.

Gray (1982) noted that the macroturbulence
dispersions, $\zeta_{\rm RT}$, are double-valued for spectral classes
G8 through K2, with weaker lines indicating larger values than
stronger lines in the same star. We have
adopted a straight average of the two sets of results
because we are going to apply the $\zeta_{\rm RT}$ results to the
line broadenings, $V_{\rm broad}$, determined from the CfA spectra
(C2003, C2007). Those velocities, in turn, were determined from a
match involving all lines, weak and strong, in a narrow wavelength
region. Finally, we replaced $\zeta_{\rm RT}$ 
values from Gray (1982) with the
newer values presented by Gray (1989).

Absolute visual magnitudes were determined for all the disk stars
using parallaxes from HIPPARCOS. Following McWilliam (1990), we
assumed that most stars had zero interstellar absorption. For the stars
for which McWilliam estimated $A_{\rm V}$ values (his Table~7), we
adopted those values. We had to exclude $\gamma^{2}$~Leo because
it does not have a measured $V$ magnitude.

We adopted uncertainties for the determinations of $V_{\rm rot}$~sin~$i$
and $\zeta_{\rm RT}$ given by Gray \& Toner (1986, 1987), and by Gray (1989). 
Gray (1982) did not provide such estimates, so we conservatively
adopted $\sigma$($V_{\rm rot}$) = $\sigma$($\zeta_{\rm RT}$) = 1.0 \kms.
For the stars from the work of Gray \& Pallavicini (1989), we
used only the ESO spectra and the lines in the $\lambda$6250 domain,
for which the errors were estimated to be $\pm 0.5$ \kms. Table~3 summarizes
the results for the 32 disk giants that satisfied the above criteria,
although as discussed below, we chose to not use all of them in
the final analyses.

\subsubsection{Halo Giants vs.\ Disk Giants}

\begin{figure}
\epsscale{0.80}
\plotone{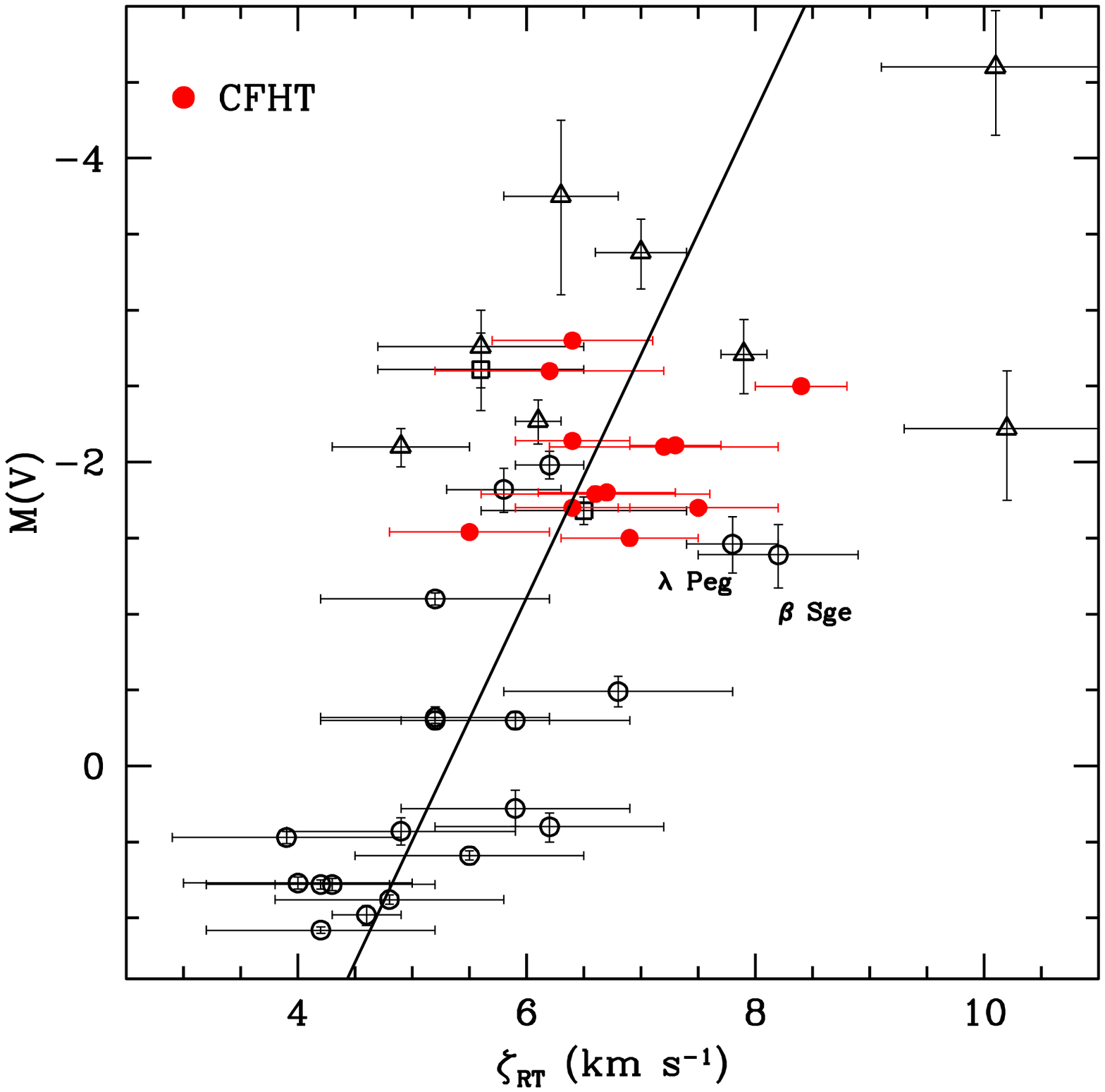}
\caption{A comparison of our derived macroturbulence dispersion
values, $\zeta_{\rm RT}$, with $M_{\rm V}$. \label{fig:zetavsmv}}
\end{figure}

\begin{figure}
\epsscale{0.80}
\plotone{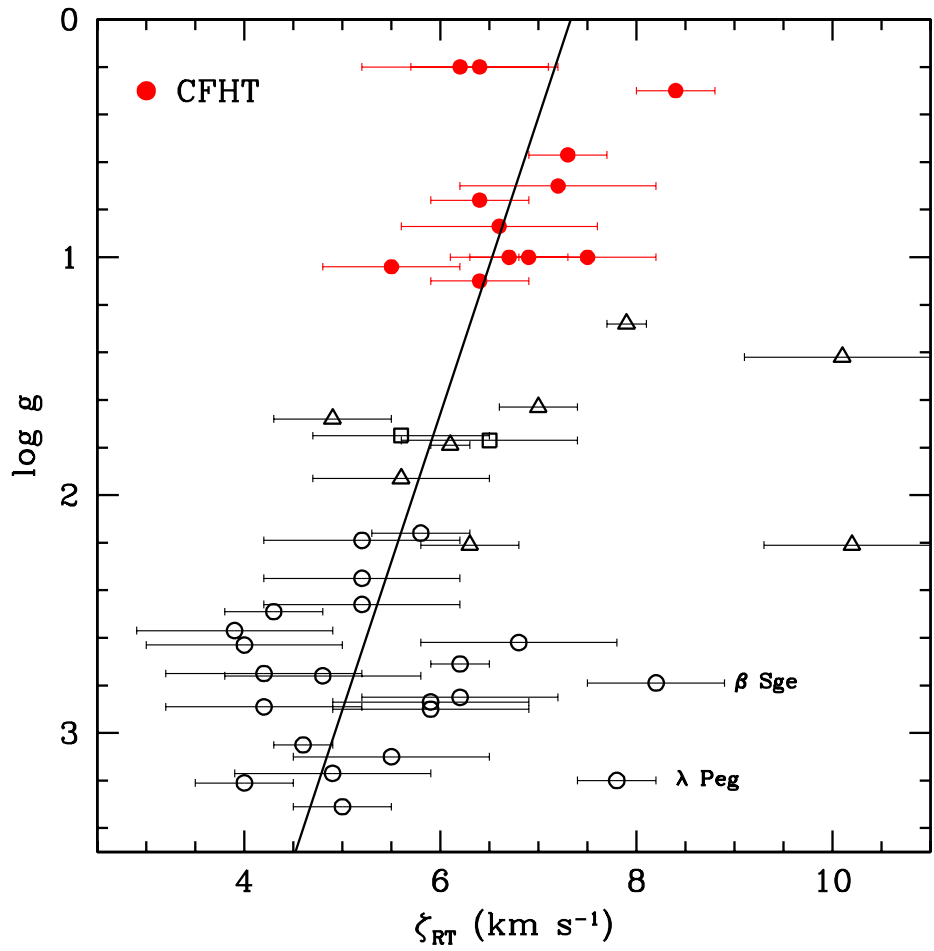}
\caption{A comparison of our derived macroturbulence dispersion
values, $\zeta_{\rm RT}$, with gravity. \label{fig:zetavsgrav}}
\end{figure}

\begin{figure}
\epsscale{0.80}
\plotone{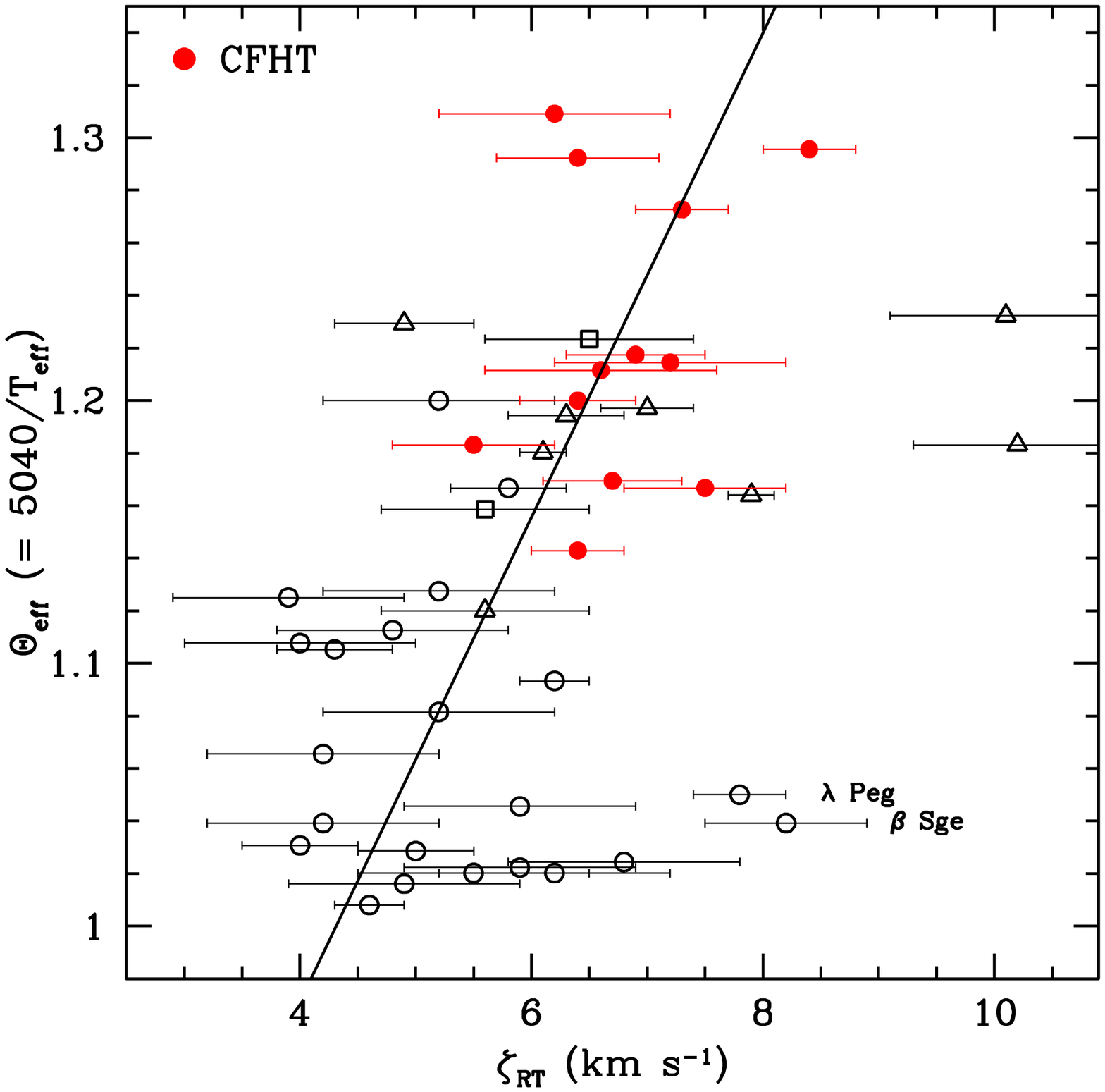}
\caption{A comparison of our derived macroturbulence dispersion
values, $\zeta_{\rm RT}$, with $\Theta_{\rm eff}$
(= 5040/$T_{\rm eff}$). \label{fig:zetavstheta}}
\end{figure}

In Figures~\ref{fig:zetavsmv} through \ref{fig:zetavstheta} we compare
the macroturbulent dispersions, $\zeta_{\rm RT}$, we have derived for
the halo giants and results for disk giants discussed above. Open
triangles are disk stars with luminosity classes Ib or II, open squares
are disk stars with luminosity class II-III, and open circles are
disk giants of luminosity class III. The filled circles represent
the halo giants observed at CFHT. Two significant results are apparent.

First, the Ib and II stars do not show a consistent trend. This should
be expected, since these stars are generally descendents of massive
stars, which do not undergo as much rotational braking while on the
main sequence, nor even during their very brief spans as red supergiants
or bright giants. 

Second, with the exceptions of $\beta$~Sge and $\lambda$~Peg, the
rest of the disk stars appear to follow the same trends as the
older halo giants. If we discard the luminosity class
Ib and II stars, as well as $\beta$~Sge and $\lambda$~Peg,
and employ weighted least squares fits, we
find:
\begin{equation}
\label{eq:zetavsmv}
\zeta_{\rm RT} = -0.62 M_{\rm V} + 5.307,
\end{equation}
with a scatter of only 0.7 \kms, and a correlation coefficient
of $-0.78$. If we consider the gravity, we find:
\begin{equation}
\label{eq:zetavsgrav}
\zeta_{\rm RT} = -0.80 \log g + 7.33.
\end{equation}
In this case the scatter is only slightly greater, 0.8 \kms,
and the correlation coefficient, $R = -0.72$, is likewise
slightly inferior. Using effective temperature as the
independent variable, we find:
\begin{equation}
\label{eq:zetavstheta}
\zeta_{\rm RT} = 10.85 T_{\rm eff} - 6.54.
\end{equation}
The scatter, 1.0 \kms, and the correlation coefficient, $-0.59$,
are a little worse than in the above two cases. While it is
true that for a fixed age and metallicity, red giants obey
monotonic relations between the three independent variables,
$M_{\rm V}$, log~$g$, and $T_{\rm eff}$, when we mix in
stars of different metallicities and ages, such correlations
fail. Consider, for example, a metal-rich disk giant with
[Fe/H= $-0.27$, [$\alpha$/Fe] = 0.0, an age of
5 Gyrs, and $T_{\rm eff}$ = 4000~K. According to the 
``Yale-Yonsei" isochrones (Yi et al.\ 2003;
Demarque et al.\ 2004), such a star has log~$g$ = +1.29
and $M_{\rm V}$ = $-0.77$. A typical halo giant with
the same temperature, but with
[Fe/H] = $-1.50$, [$\alpha$/Fe] = +0.3, and an age of 10 Gyrs
has log~$g$ = +0.50 and $M_{\rm V}$ = $-2.40$. 
Figures~\ref{fig:zetavsmv}-\ref{fig:zetavstheta} indicate that
whether one considers $M_{\rm V}$, log~$g$, or $T_{\rm eff}$
as the independent variable, metallicity differences do not
seem to have significant effects on the derived 
macroturbulence relations.

\section{REVISED ROTATIONAL VELOCITIES FOR THE FULL CFA SAMPLE}

\subsection{Red Horizontal Branch Stars}

\subsubsection{Macroturbulence}

Equation~\ref{eq:vbroad} offers us an opportunity to estimate
stellar rotational velocities for the stars observed at CfA
but not at CFHT, \underline{if} we have some knowledge of
$\zeta_{\rm RT}$. In the case of the RHB stars, this turns out
to be a vexing problem.

\begin{figure}
\epsscale{0.80}
\plotone{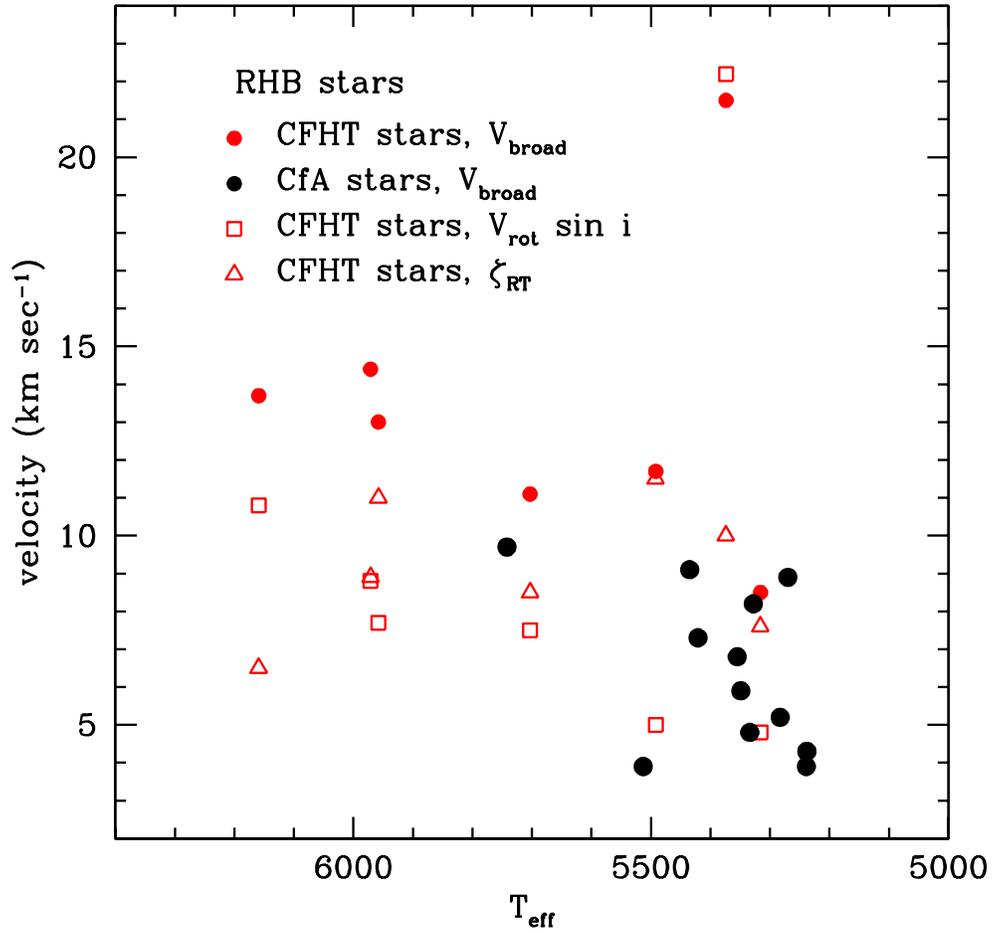}
\caption{The behavior of rotational velocity, $V_{\rm rot}$~sin~$i$
(open squares), macroturbulence dispersion, $\zeta_{\rm RT}$
(open triangles), and line broadening derived by
C2003 and C2007 (filled circles), as a function of $T_{\rm eff}$. 
Red symbols refer to the
seven red horizontal branch stars with results given in Table~2.
The black filled circles represent the $V_{\rm broad}$
values for the thirteen other RHB stars studied
by C2003 and C2007.
\label{fig:rhbvbroadvsteff} }
\end{figure}

C2007 observed that macroturbulence might explain the 
monotonic rise in line broadening with
increasing effective temperature seen in their sample of 20 field
RHB stars.
The suggestion relied primarily on the finding by Gray \& Toner (1986)
that $\zeta_{\rm RT}$ is a significant function of spectral type
(i.\ e., temperature) within individual luminosity classes. 
Figure~\ref{fig:rhbvbroadvsteff} shows a modified version of
Figure~12 from C2007. Filled circles represent the line broadenings
reported by C2003 and C2007, with the red ones representing stars
for which we have derived rotational and macroturbulent velocities
(Table~2). Open squares depict $V_{\rm rot}$~sin~$i$, and open
triangles show $\zeta_{\rm RT}$. If 
we focus on the stars in red (from Table~\ref{tab2}), we note
the near-constancy of the macroturbulent dispersion, $\zeta_{\rm RT}$,
from $T_{\rm eff}$ $\approx\ 5300$~K to 6200~K.
The value appears to be constant, unlike that predicted 
by Gray \& Toner (1986), although the temperature
spread is not large, and the sample size is small.

There is a more fundamental concern. The mean value of $\zeta_{\rm RT}$
for the seven RHB stars observed at CFHT is $9.1 \pm 0.7$ \kms.
If Equation~\ref{eq:vbroad} is correct, then none of the other
thirteen RHB stars observed only at CfA should have $V_{\rm broad}$
values smaller than that, but Table~\ref{tab4} contains several
values near 4 \kms. Figure~\ref{fig:rhbvbroadvsteff} shows that the seven
stars observed at CFHT are a bit hotter than the thirteen stars
observed only at CfA. If $\zeta_{\rm RT}$ depends on $T_{\rm eff}$,
as expected, could that explain the disagreement? We answer that
question by considering only the stars in the narrow temperature
range of 5200~K to 5600~K. The three stars observed at CFHT have
$<\zeta_{RT}> = 9.7 \pm 1.1$ \kms ($\sigma$ = 2.0 \kms). The twelve
stars observed only at CfA have $<V_{\rm broad}> = 6.1 \pm 0.5$ \kms,
significantly lower than Equation~\ref{eq:vbroad} would predict.

Are the CfA results reliable
at such low $V_{\rm broad}$ values? Recall that the CfA instrumental resolution
is about 8.5 \kms. Six of the twenty
RHB stars studied by C2003 and C2007 were also studied by
Behr (2003), using higher resolution and higher-S/N spectra,
and the mean difference in derived line broadenings
for those six stars is only $+0.8 \pm 0.4$ \kms, $\sigma = 1.0$ \kms.  
For the two stars common to both sets of studies and with the lowest
line broadening, Behr (2003) reports values of 6.6 and 5.4 \kms\ for
BD+11~2998 and BD+9~3223, while C2003  found 6.8 and 4.8 \kms,
respectively. The CfA $V_{\rm broad}$ values appear to be reliable.

The observational bias of our program may be partly to blame.
Most of the stars we observed were selected for study because
of their larger-than-average $V_{\rm broad}$ values. If that also
means that we have selected stars with larger-than-average
$\zeta_{\rm RT}$ values, then we might understand the discrepancy
between the CFHT results for $\zeta_{\rm RT}$ compared to the
smaller values of $V_{\rm broad}$ for the remaining RHB stars.
But if that is the explanation, then the validity of
Equation~\ref{eq:vbroad} must be questioned because in that
case because $\zeta_{\rm RT}$ would itself be highly variable
even within a narrow range of temperature.

In the absence of a simple explanation, we're forced to conclude
that whereas the macroturbulence of red giants appears well-behaved
(see Figures~\ref{fig:zetavsmv} to \ref{fig:zetavstheta}), that is not
the case for RHB stars. This might be due to short-term variability
of the phenomenon, for example. It might also be due to differences in
evolutionary state that we cannot readily discern in field stars.
For example, some of the field stars may have begun core helium
burning with a surface temperature cooler than the instability
strip; what we would call zero-age red horizontal branch stars.
Other stars may have begun core helium burning within the
instability strip (as RR~Lyrae variables) or even hotter (BHB
stars). There is a trend among globular clusters' horizontal
branches such that the more metal-poor clusters tend to have
horizontal branches populated mostly by BHB stars while the more
metal-rich clusters favor RHB stars. Most BHB stars eventually
evolve back across the H-R diagram en route to the asymptotic
giant branch, so more highly evolved stars now in the RHB domain
might be distinguishable statistically by lower metallicities.
Let us consider the thirteen stars observed at CfA for which
we do not have CFHT spectra. The seven stars with $V_{\rm broad} \leq\ 6$ \kms\
have $<$[Fe/H]$>$ = $-1.99 \pm 0.23$ ($\sigma$ = 0.62), while
the six other stars, with $6.8 \leq\ V_{\rm broad} \leq\ 9.7$ \kms,
have $<$[Fe/H]$>$ = $-1.61 \pm 0.19$ ($\sigma$ = 0.47). There
is marginal evidence for the stars with smaller broadening values
being more metal-poor (and more likely to have evolved away from
the BHB and now be more luminous and slightly larger in radius
than zero-age RHB stars). The argument fails, however, when we
consider the seven RHB stars in Table~\ref{tab2} with large rotational velocities
and large $\zeta_{\rm RT}$, since $<$[Fe/H]$>$ = $-2.02 \pm 0.15$ 
($\sigma$ = 0.39). 

We believe that the simplest interpretation at this point is
to admit that the RHB stars do not share a well-defined
relationship between macroturbulent dispersion and effective
temperature.

\subsubsection{Rotational Velocities}

\begin{figure}
\epsscale{0.80}
\plotone{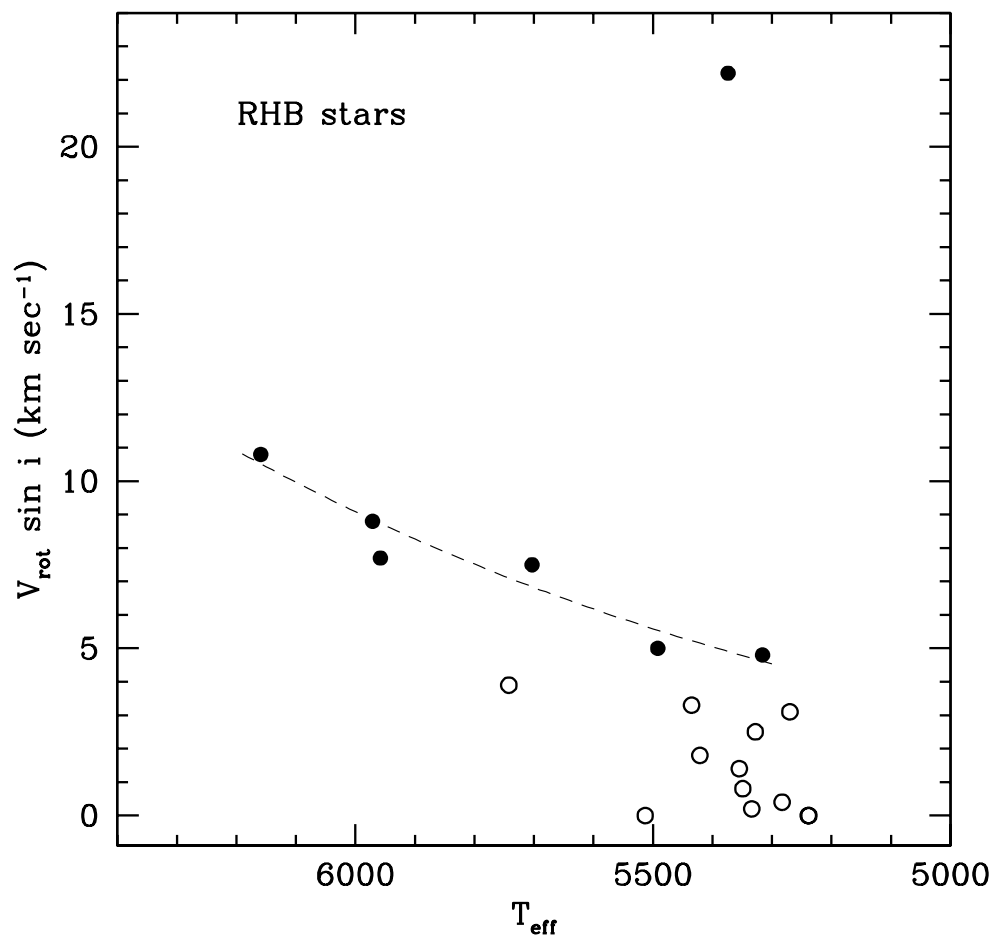}
\caption{The behavior of rotational velocity, $V_{\rm rot}$~sin~$i$
for red horizontal branch stars as a function of $T_{\rm eff}$.
Filled circles are values from Table~\ref{tab2} while open
circles depict results from Equation~\ref{eq:vbroadcorr} applied
to the other 13 RHB stars observed at CfA. The dashed line
represents the results of a log-log relationship derived using
six of the CFHT stars, $V_{\rm rot}$~sin~$i \propto T_{\rm eff}^{5.6}$. 
We excluded HD~195636.
\label{fig:rhbvrotvsteff} }
\end{figure}

While we are not confident in our ability to ``remove" the
contribution of macroturbulence to the values of $V_{\rm broad,CfA}$
using Equation~\ref{eq:vbroad}, we may still estimate rotational
velocities for the other thirteen RHB stars using
Equation~\ref{eq:vbroadcorr}. The basic stellar parameters for
these stars, taken from C2003 and C2007, are given in 
Table~\ref{tab4}, which includes the results from 
the application of Equation~\ref{eq:vbroadcorr}.
Figure~\ref{fig:rhbvrotvsteff} shows the results, with filled circles
representing $V_{\rm rot}$~sin~$i$ values from Table~\ref{tab2}
and open circles representing values deriving employing 
Equation~\ref{eq:vbroadcorr}. As expected, the stars observed at
CFHT have larger rotational velocities, due to our bias toward
stars with larger line broadening. Excluding the anomalous
star HD~195636, the average rotational velocity for the remaining
nineteen RHB stars is $3.3 \pm 0.8$ \kms\ ($\sigma$ = 0.8 \kms).

There is an apparent trend in the upper limits of $V_{\rm rot}$~sin~$i$
as a function of $T_{\rm eff}$, with lower values at cooler
temperatures (again, we neglect HD~195636). We do expect
some sort of trend, \underline{if} all six stars have identical
amounts of rotational angular momentum and have comparable values of
rotational axis inclinations. In that particular case, 
since $L \propto\ R^{2}T_{\rm eff}^{4}$ and the rotational
angular momentum $J \propto\ MR^{2}$, then for equal masses,
$V_{\rm rot}$~sin~$i \propto\ T_{\rm eff}^{2}$. Using a
log-log calculation we found that the six RHB stars 
with $V_{\rm rot}$~sin~$i$ values in Table~\ref{tab2} define
$V_{\rm rot}$~sin~$i \propto\ T_{\rm eff}^{5.6}$, which is
shown as the dashed line in Figure~\ref{fig:rhbvrotvsteff}. 
However, a straight line, $V_{\rm rot}$~sin~$i \propto T_{\rm eff}$, 
provides almost as good a
representation of the limited data, so no definitive
conclusions may be drawn.

\subsubsection{Comparison of Rotation in Blue and Red Horizontal Branch Stars}

Finally, we can infer one interesting result, somewhat related
to the issue raised above regarding rotational angular momentum
as a function of temperature. Kinman et al.\ (2000) provided
estimates of $V_{\rm rot}$~sin~$i$ for 30 blue horizontal branch (BHB)
stars. The question we ask is whether the RHB stars we have
studied show as large a range in rotational velocities 
as BHB stars, when allowance is
made for the different stellar radii and moments of inertia. 
Because the method employed by Kinman et al.\ (2000) was based
on the full width at half maximum of the $\lambda$4481 Mg~II line,
and their spectral resolving power was only about 15 \kms, we
ask what fractions of their BHB sample were found to have $V_{\rm rot}$~sin~$i$
larger than 15 and 20 \kms, and then ask
what fractions of our RHB have similar rotational velocities when
allowances are made for the different stellar radii.

We consider first the BHB sample. We derived stellar radii by
converting the log~$g$ values into radii, assuming a stellar
mass of 0.7~$M_{\odot}$. This is a somewhat smaller value
than the mass we adopted for the RHB stars, 0.8~$M_{\odot}$,
and leads to a mean stellar radius of 3.3~$R_{\odot}$ (excluding
BD+32~2188
because the radius derived
from its log~$g$ value, 11.8~$R_{\odot}$, suggests
it is not a BHB star). 
Six of the remaining 29 BHB stars (21\%)
have estimated $V_{\rm rot}$~sin~$i \geq\ 20$ \kms, and 10 of
them (34\%) have $V_{\rm rot}$~sin~$i \geq\ 15$ \kms. To compare
to the RHB sample of Table~2, we must reduce the $V_{\rm rot}$~sin~$i$
limits by the ratio of the stellar radii, (3.3/7.5), so the 20 \kms\ limit
becomes 8.8 \kms\ and the 15 \kms\ limit becomes 6.6 \kms.
Any difference in mass between the more massive RHB stars and
the lower mass BHB stars would lower these limits further.
Of the 7 RHB stars in Table~2, three (43\%) 
have $V_{\rm rot}$~sin~$i \geq\ 8.8$ \kms,
and five (71\%) 
have $V_{\rm rot}$~sin~$i \geq\ 6.6$ \kms. This small
sample would suggest that the RHB
stars show a higher fraction of relatively rapid rotators than
do the BHB stars, but we reiterate that the stars listed in Table~\ref{tab2}
were selected on the basis of large $V_{\rm broad}$ values, so we must,
in fact, include the thirteen other RHB stars, for which we have only
estimated $V_{\rm rot}$~sin~$i$. The mean radius of the 20 RHB
stars is 8.0~$R_{\odot}$, so the 20 and 15 \kms\ limits 
for BHB stars are now 8.3 and 6.2 \kms, respectively, again
neglecting mass differences.
None of the thirteen RHB stars have
$V_{\rm rot}$~sin~$i$ values larger than 4 \kms, so the percentage
of RHB stars with $V_{\rm rot}$~sin~$i$ greater than 8.3 and 6.2 \kms\
drops to 15\% and 25\%, compared to the BHB stars' values (scaled for
radius) of 21\% and 34\%. 

RHB stars may show another similarity to BHB stars in the
distribution of their rotational velocities. Peterson, Rood,
\& Crocker (1995) found a bimodal distribution of rotational
velocities in the metal-poor globular cluster M13, which has
a predominantly blue horizontal branch. Behr (2003) noted an
apparent bimodal distribution in the rotational velocities
of metal-poor field BHB stars as well. Based only on one
star, HD~195636, there might also be a ``bimodality" in the
rotational velocity distribution of RHB stars.

Unfortunately, the small sample sizes
prohibit any firm conclusions, except that there is no compelling
evidence that the rotational angular momentum of RHB stars is
any smaller than that of BHB stars.

\subsection {Rotation of Red Giant Branch Stars}

\subsubsection{CFHT Data}

\begin{figure}
\epsscale{0.80}
\plotone{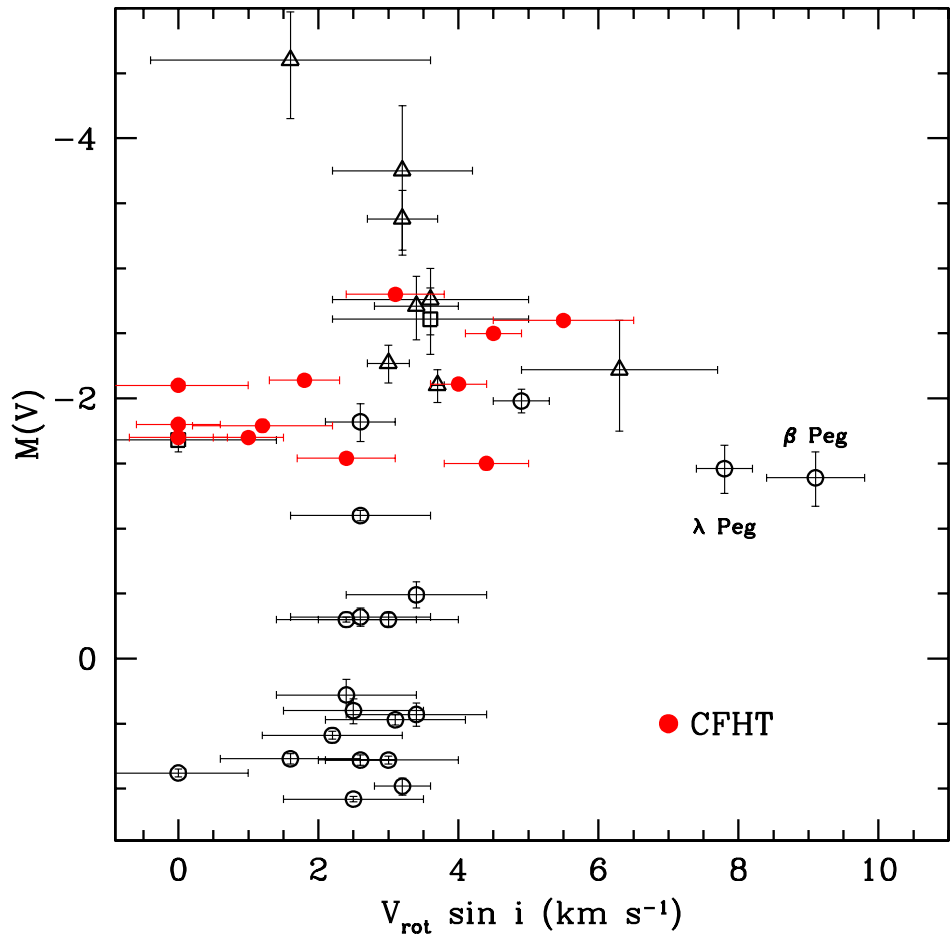}
\caption{The $V_{\rm rot}$~sin~$i$ values for red giants
derived from Fourier transform methods as a function of
luminosity. Stars from Table~2 are plotted as filled
red circles. \label{fig:giantsvrotvsmv}}
\end{figure}

Figure~\ref{fig:giantsvrotvsmv} shows the rotational velocities
derived from the Fourier transform analyses
as a function of visual absolute magnitude for the stars in
Tables~2 and 3, including giants, bright giants, and supergiants.

Consider the spread in $V_{\rm rot}$~sin~$i$ values for the
disk giants. 
(As before, we exclude
$\beta$~Sge and $\lambda$~Peg.) Sixteen of the eighteen disk giants
have $V_{\rm rot}$~sin~$i$ values between 1.5 and 3.5 \kms, a
quite narrow range. The halo giants appear to be more evenly
spread out in $V_{\rm rot}$~sin~$i$, from about 0 to 5.5 \kms.
Both distributions are puzzling. For example,
if we assume that all the halo RGB stars have
similar rotational velocities, the distribution in $V_{\rm rot}$~sin~$i$ is
not consistent with the expected distribution of viewing angles.
If $V_{\rm rot}$ = 5.0 \kms, we expect to find almost
two thirds of the stars with $V_{\rm rot}$~sin~$i$ $\geq\ 3.0$ \kms,
but only about one sixth of the stars would have $V_{\rm rot}$~sin~$i$ $\leq\ 2.0$
\kms. Instead we find five out of the twelve (42\%) with the higher
rotational velocities, and half with the lower velocities. While
the statistics are weakened by the small sample size, it appears
that more is at work here than just geometry.

\subsubsection{CfA Data}

Equations~\ref{eq:vbroad} and \ref{eq:vbroadcorr} allow us,
in principle, to estimate $V_{\rm rot}$~sin~$i$ for all of
the 116 red giants in the CfA program (C2003, C2007). In the
case of Equation~\ref{eq:vbroad}, we must adopt some
representation of $\zeta_{\rm RT}$, but Equations~\ref{eq:zetavsmv}
through \ref{eq:zetavstheta} appear to be well-behaved. 

For several reasons, we prefer to explore
the behavior of the rotation of this larger sample of metal-poor field
red giants in a slightly different fashion. As noted already, projection
effects compromise the results for individual stars. 
Figure~\ref{fig:zetavsmv} shows that the relation between $\zeta_{\rm RT}$
and $M_{\rm V}$ appears well-behaved, but there is scatter among the
individual stars, so the correction is best treated in a statistical
fashion. Finally, while Figure~9 of C2007 shows that the $V_{\rm broad}$
values determined by C2003 and C2007 are good measures of line
broadening down to values as low as 3 \kms\ (despite an instrumental
resolution of 8.5 \kms), uncertainty remains for each individual star.
We conclude that it is best to approach the behavior of
rotational velocities derived in the above fashion in a statistical
manner. 

We began our analysis by removing three stars from the CfA sample,
all of them binaries with short periods and, hence relatively
close separations, where tidal effects of a companion may induce or inhibit
the rotation of the primary star. All three stars have unusually
small orbital eccentricities, consistent with tidal interactions.
With BD+30~2034, BD+18~2890, and CD$-37$~14010 removed from the sample,
we sorted the remaining 113 metal-poor red giants by $M_{\rm V}$
and averaged the results within six bins, chosen
simply so that the five more luminous bins
have equal numbers of stars (20), while the lowest
luminosity bin has 13.  Table~5 contains the
results, including $<M_{\rm V}>$, $<R/R_{\odot}>$, and 
$<V_{\rm broad}>$. The Table also includes the error, $\sigma$, and the
error of the mean, $\sigma_{\mu}$, for each quantity. The last two
columns of Table~5 include the resultant mean rotational velocity,
obtained using $<V_{\rm broad}>$ and Equations~\ref{eq:zetavsmv}
and \ref{eq:vbroad}, and Equation~\ref{eq:vbroadcorr}, respectively. 
In cases where the macroturbulence value
exceeded $V_{\rm broad}$, we set $V_{\rm rot}$~sin~$i$ to zero.
Note that the results for the most luminous stars obtained 
using Equation~\ref{eq:vbroadcorr} agree well with
the mean results for the twelve stars observed at
CFHT ($<M_{\rm V}> = -2.0$; $<V_{\rm rot}$~sin~$i>$ = 2.3 \kms).

\begin{figure}
\epsscale{0.80}
\plotone{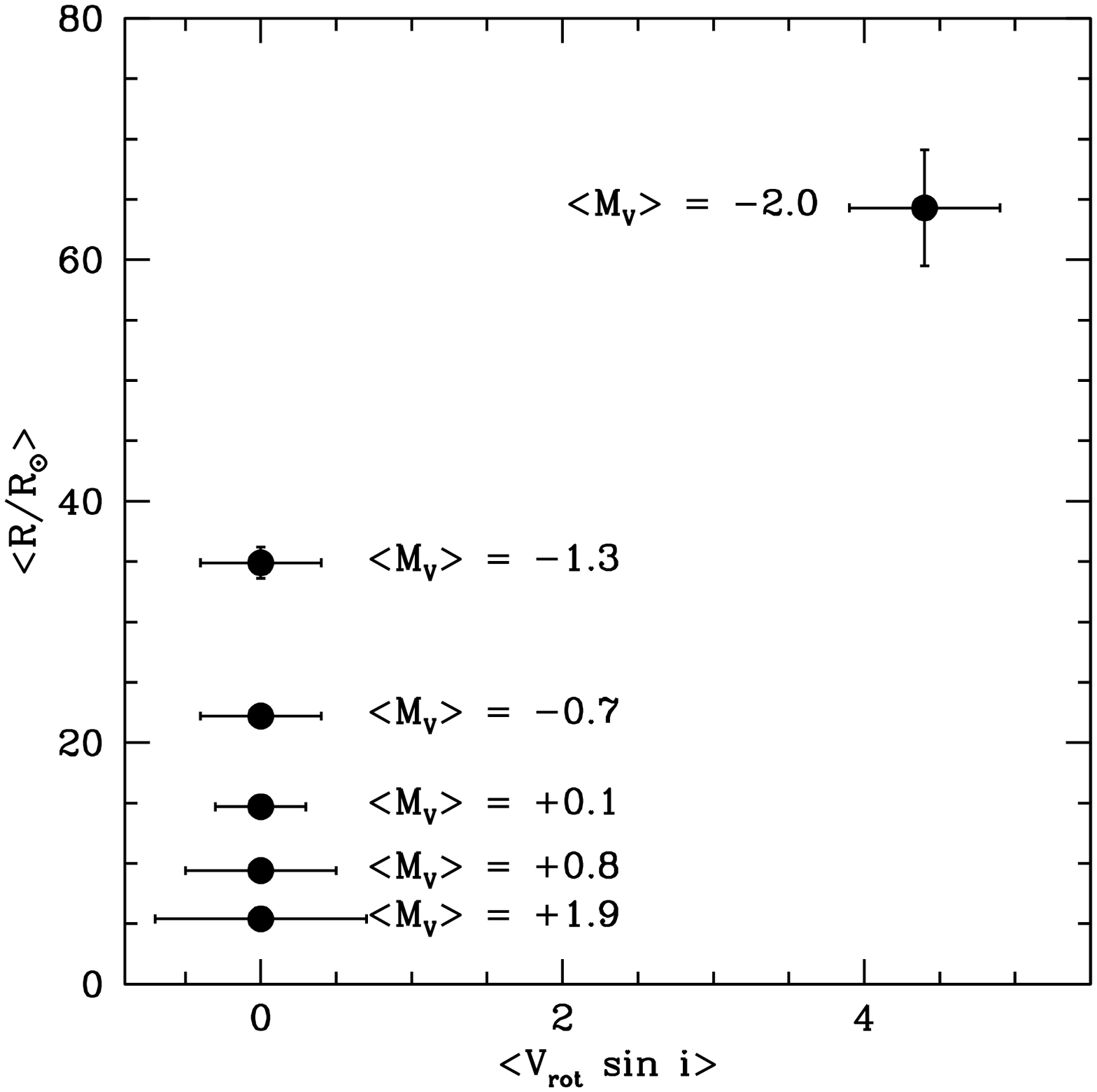}
\caption{The binned results for stellar radii and
stellar rotational values for metal-poor red giants from Table~5 are compared.
For stars in five of the $M_{\rm V}$ bins, the estimated 
values of $\zeta_{\rm RT}$ equal or exceed the average
$V_{\rm broad}$ values, and the rotational velocities
are set to zero. We show also the mean $M_{\rm V}$ values
for each of the six bins. These results were computed by
removing the effects of $\zeta_{\rm RT}$ on total line broadening
using Equation~\ref{eq:vbroad}.
\label{fig:allrgbvrotmeans}}
\end{figure}

Figures~\ref{fig:allrgbvrotmeans} and \ref{fig:revallrgbvrotmeans} 
show the results. In the former case we have employed
Equations~\ref{eq:vbroad} and \ref{eq:zetavsmv}, while in the
second case we used Equation~\ref{eq:vbroadcorr}. 
We have plotted the mean rotational velocities as a function
of stellar radius, which we assume, to first order, tracks 
the rotational angular momentum
since all the stars should have very similar masses and mass
distributions. The two binned samples show zero or near-zero
rotation at all radii (and luminosities) except for the largest radii.
This contradicts any stellar evolution that
includes conservation of angular momentum, so
either the surface has acquired extra rotation from internal
sources or from external ones (such as absorption of a large
planet), or the line broadening to which we refer as rotation
is something else. 

\subsubsection{Discussion}

\begin{figure}
\epsscale{0.80}
\plotone{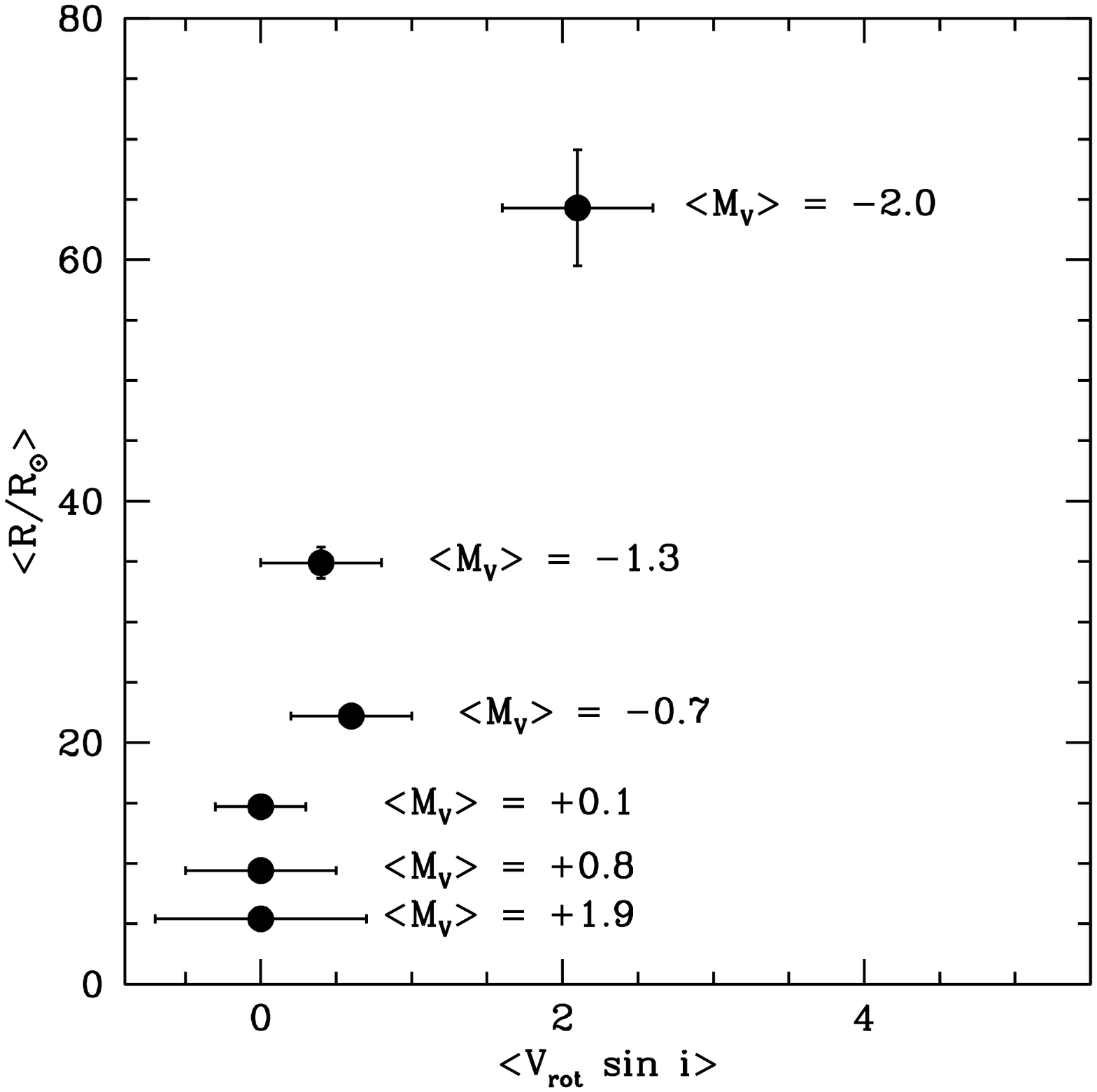}
\caption{The binned results for stellar radii and
stellar rotational values for metal-poor red giants from Table~5 are compared.
For stars in five of the $M_{\rm V}$ bins, the estimated 
values of $\zeta_{\rm RT}$ equal or exceed the average
$V_{\rm broad}$ values, and the rotational velocities
are set to zero. We show also the mean $M_{\rm V}$ values
for each of the six bins. These results were computed 
using Equation~\ref{eq:vbroadcorr}, which does not make use
of $\zeta_{\rm RT}$.
\label{fig:revallrgbvrotmeans}}
\end{figure}

The results of Figures~\ref{fig:allrgbvrotmeans} and \ref{fig:revallrgbvrotmeans}
are hard to understand. C2003 and C2007 discussed the concept of transport
of internal angular momentum to the surface layers, but that should
appear at the radius and luminosity when the convection zone reaches its
deepest penetration. Conceivably, Figure~\ref{fig:revallrgbvrotmeans}
is showing such an effect at $M_{\rm V} \approx\ -0.7$. But that still
does not explain the rotation seen (only) among the largest radii,
most luminous stars. The planet search effort of Sozzetti et al.\ (2006)
suggests that absorption of planets is not sufficiently common to
explain the results, either. 
The appearance of additional (admittedly small) line broadening
among only the most luminous metal-poor red giants invites a
comparison with two other phenomena that likewise appear favored
among such stars: velocity jitter and mass loss.

C2003 and C2007 summarized the appearance of jitter in metal-poor
red giants. Jitter becomes increasingly common, with typical
velocity variations of about 2 \kms\ for $M_{\rm V} < -1.4$. 
For the stars studied as part of this program,
a careful consideration of the results
in Tables~2 and 3 shows that 
all five of the stars with $V_{\rm rot}$~sin~$i \geq\ 3.0$ \kms\
show velocity jitter, defined in C2003 and C2007 as stars with
P($\chi^{2}$) $\leq\ 10^{-6}$. Of the 12 red giants in Tables~2 and 3,
the mean rotational velocity for the 8 stars showing velocity jitter
is $2.9 \pm 0.7$ \kms, compared to $1.1 \pm 0.5$ \kms\ for the other
4 stars without detectable velocity jitter. 

\begin{figure}
\epsscale{0.80}
\plotone{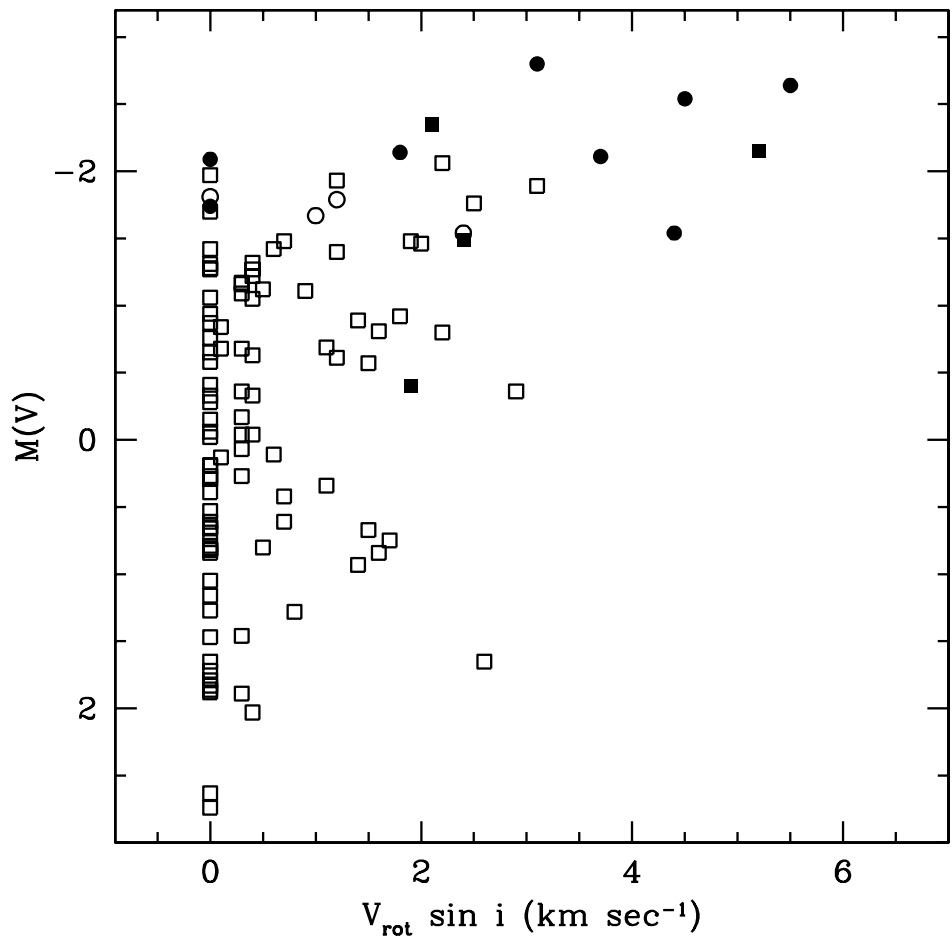}
\caption{The individual results for stellar rotational
velocities from Table~\ref{tab2} (circles) or use of 
Equation~\ref{eq:vbroadcorr} (squares) are
plotted against $M_{\rm V}$.
Filled circles and squares represent stars that display jitter,
defined as P($\chi^{2}$) $\leq\ 10^{-6}$, exclusive
of orbital motion. Open circles and squares are stars that have
not been found to manifest such random velocity
variability. \label{fig:jitter}}
\end{figure}

In Figure~\ref{fig:jitter} we display the results for individual
stars that went into the bins of Table~5 and Figure~\ref{fig:revallrgbvrotmeans}.
Stars displaying velocity jitter are shown as filled circles.
The highest luminosity bin in Figure~\ref{fig:revallrgbvrotmeans}
contains the twenty metal-poor red giants with $M_{\rm V} \leq\ -1.5$.
Inspection of Figure~\ref{fig:jitter} shows that 11 of the 17 stars in that
bin with $V_{\rm rot}$~sin~$i \geq\ 3.0$ \kms\ show jitter. Clearly
velocity jitter is related in some way to the excess line broadening.
Does more rapid rotation among the most luminous
stars induce jitter? Does velocity jitter masquerade as 
rotational broadening in some fashion? Or are both phenomena,
excess line broadening and velocity jitter, symptoms of the
same cause?

Mass loss may be another key to the puzzle. Smith et al.\ (1992),
Dupree \& Smith (1995), and Cacciari et al.\ (2004) studied
Ca~II K$_{2}$ line profiles in metal-poor giants, finding that
for $M_{\rm V} < -1.7$, the line emission is weaker on the
violet side than on the red side, so that the ratio of violet to red
emission line flux, V/R, is less than unity for the most metal-poor stars.
That ratio indicates mass outflow, and it is therefore interesting
that we have three phenomena, excess line broadening, velocity
jitter, and mass outflow, that appear only at the high
luminosities found near the red giant branch tip for metal-poor
stars. Surely Occam's Razor suggests a common cause.

\begin{figure}
\epsscale{0.80}
\plotone{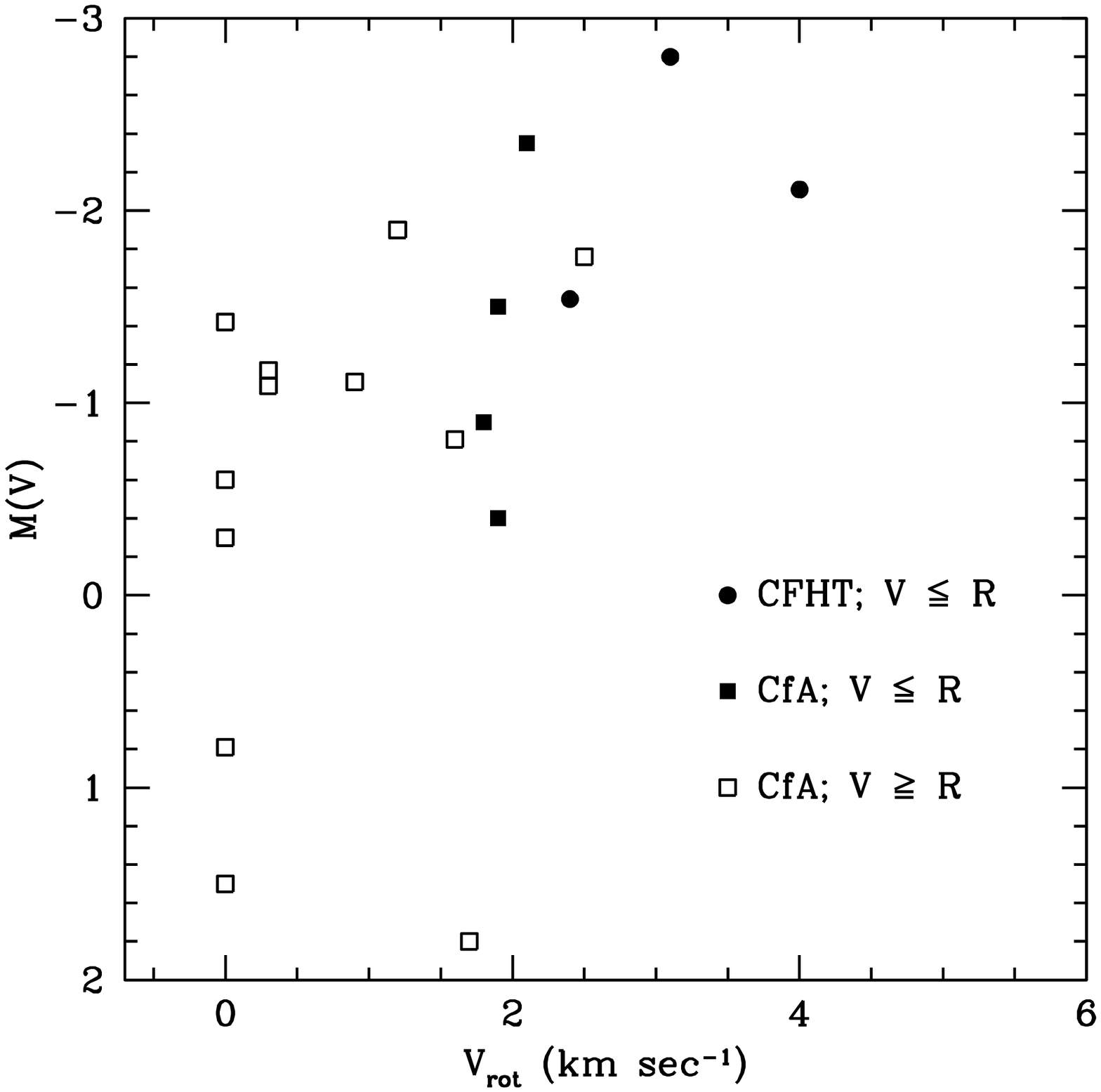}
\caption{Rotational velocities of red giant stars obtained using
CFHT spectra (filled circles), and CfA spectra, corrected using
Equation~\ref{eq:vbroadcorr} (open circles). 
Filled circles and squares signify
stars with greater emission on the red side of Mg~II or Ca~II 
lines, indicative of mass outflow. Open squares indicate 
stars with greater flux on the
violet side of the emission lines.
\label{fig:vrotvsmv}}
\end{figure}

Recently, Dupree et al.\ (2007) discussed the mass outflow on
the basis of $\lambda$2800 Mg II absorption profiles in 
metal-poor stars, finding that the V/R asymmetry indicative
of mass loss appears at somewhat fainter luminosities 
than for the Ca~II K$_{2}$ asymmetry, or that of H$\alpha$.
In Figure~\ref{fig:vrotvsmv} we show how Ca~II and Mg~II
V/R ratios vary among stars observed by C2003, C2007, and
in this paper. As in the case of jitter, it is clear that
high luminosity favors both excess line broadening as well
as mass loss. Further, as Dupree et al.\ (2007)
stressed, the V/R ratios in some stars are variable, suggesting
that mass loss is episodic. Thus some of the stars in
Figure~\ref{fig:vrotvsmv} with V$>$R may be temporarily
stable in terms of mass loss.

Dupree et al.\ (2007) also noted that the mass loss decreases
at lower metallicities, although the trend is weak.
Mass loss remains significant in halo giants, despite the lower
metallicities and especially the greater ages of halo giants
compared to disk giants. This is unexpected for mass loss
driven by magnetic processes, but might be explicable in
terms of acoustic shock wave heating of the chromospheres
of metal-poor giants (Cuntz et al.\ 1994). The models
employed in that paper suggested velocity variations of
a few \kms\ with periods probably shorter than pulsation,
so this might be a mechanism for producing velocity
jitter as well as additional line broadening.

Another explanation
is that pulsation may drive heating and mass loss (Smith \& Dupree 1988).
Pulsation is an attractive model since it could also
explain velocity jitter and possibly extra line
broadening due to the accelerations present in a pulsating
atmosphere. C2003 drew attention to the apparent periodicity in
the velocity jitter of HD~3008 (172 days; amplitude 1.55 \kms)
and BD+22~2411 (186 days; amplitude 0.96 \kms). The periods
are long compared to known long-period variables in metal-poor
clusters, but the periodicity is certainly suggestive. It would
be worthwhile to explore more carefully the line broadening,
radial velocity, and mass loss together
in metal-poor luminous giants on a variety of timescales.

\section{CONCLUSIONS}

We have obtained high-resolution, high-S/N spectra for 12 metal-poor
field RGB stars and 7 metal-poor field RHB stars. Fourier transform
analyses have yielded good estimates for the macroturbulence
dispersion, $\zeta_{\rm RT}$, and the rotation velocity,
$V_{\rm rot}$~sin~$i$, for all the stars. We obtained consistent
results for HD~29574, which was observed during both observing
runs, and for $\eta$~Ser, which had been studied previously by
Gray \& Pallavicini (1989). It is good to recall that we selected
our stars from the C2003 and C2007 samples of 116 RGB stars and
20 RHB stars, and with a bias toward stars with larger line
broadening values (referred to as rotational velocities by
C2003). The RGB stars appear to show very similar
macroturbulence behavior as a function of luminosity, gravity,
and temperature as do disk giants. The twelve RGB stars studied
here, however, show a larger range in rotational velocities,
with half showing values of 2.0 \kms\ or less, and five having
values in excess of 3.0 \kms. For the seven field RHB stars,
we confirm the rapid rotation of HD~195636 discovered by
Preston (1997). When allowance is made for the star's larger
radius compared to BHB stars, its rotational angular momentum
is comparable to the largest seen in field BHB stars. 

We have explored the use of two empirical methods, neither
justified physically, to attempt to exploit the more extensive
data on line broadening from C2003 and C2007. The derived
rotational velocities of the 13 field RHB stars are, as expected
from the CfA results, modest. To compare our results with the
much lower resolution BHB data from Kinman et al.\ (2000), we
consider the percentages of BHB stars whose rotation rates
are comparable to the spectral resolution, and then make
allowances for the differences in radii between the BHB stars
and our sample of RHB stars. We find that the RHB stars have
fewer rapidly rotating stars than does the BHB sample studied
by Kinman et al.\ (2000). 

Both algorithms were applied to binned samples of the field
RGB stars, so that statistical fluctuations would be diminished.
All but the bin containing the most luminous stars
($M_{\rm V} \leq\ -1.5$) showed nearly zero mean rotation,
as expected for the large radii. 
It is clear that the most luminous
metal-poor field RGB stars show enhanced rotation or some
other source of line broadening, as
found initially by C2003. Unlike the results from C2003,
however, the line broadening is relatively modest, on
average being 2 \kms\ or 4 \kms, depending on
which algorithm is employed. Our CFHT observations did
not extend to the lower luminosities, but are consistent
with this result. The twelve RGB stars studied had
$<M_{\rm rot}>$ = $-2.0$ and 
$<V_{\rm rot}$~sin~$i>$ = 2.3 \kms. 

We draw attention to the fact that the transition in luminosity
between negligible and significent rotation 
occurs at about the same $M_{\rm V}$ value as the appearance
of velocity jitter and mass loss. This is highly suggestive of
a common underlying physical origin, which may be
a sign of shock waves and acoustic heating of chromospheres
(Cuntz et al.\ 1994) or pulsation (Smith \& Dupree 1988). 
We recommend a dedicated monitoring program of a few
key stars to compare timescales and the presence of
line broadening, velocity variations, and mass loss.

We acknowledge financial support from the National Science Foundation
to the University of North Carolina through grant AST0305431 and
to Bowling Green State University through grant AST0307340.

\clearpage

\begin{deluxetable}{lrrrrrrrl}
\tablewidth{0pc}
\tablenum{1}
\footnotesize
\tablecaption{Observational Data \label{tab1}}
\tablehead{
\colhead{Star} &
\colhead{$\lambda$} &
\colhead{HJD$-$2,450,000} &
\colhead{$V$} &
\colhead{Exp} &
\colhead{S/N\tablenotemark{a}} &
\colhead{$V_{\rm rad}$} &
\colhead{$\sigma$} &
\colhead{Comments} }
\startdata
HD~3008    & 6150 & 3366.7716 & 9.70 & 40 & 175 & $-81.83$ & 0.34 & RGB; CM~Cet; jitter \nl
BD$-$18~271 & 6150 & 4015.8939  & 9.85 & 90 & 145 & $-210.54$ & 0.19 & RGB; jitter \nl
CD$-36$~1052 & 5430 & 3366.8083 & 10.00 & 70 & 150 & $+304.36$ & 0.32 & RHB \nl
HD~23798   & 6150 & 3366.8681 & 8.32 & 25 & 210 & $+88.83$ & 0.54 & RGB \nl
HD~25532   & 5430 & 4016.9720 & 8.24 & 80 & 175 & $-111.88$ & 0.34 & RHB \nl
BD+6~648   & 6150 & 4015.1177 & 9.09 & 270 & 285 & $-142.41$ & 0.31 & RGB \nl
HD~29574   & 6150 & 3366.8934 & 8.38 & 60 & 155 & $+17.86$ & 0.37 & RGB; HP~Eri; jitter \nl
           & 6150 & 4015.0694 &      & 35 & 250 & $+17.67$ & 0.48 & \nl
HD~82590   & 5430 & 3366.9454 & 9.42 & 75 & 180 & $+214.31$ & 0.37 & RHB; NSV~4526 \nl
BD+22~2411 & 6150 & 3367.0135 & 9.95 & 90 & 160 & $+35.05$ & 0.26 & RGB; jitter \nl
HD~106373  & 5430 & 3367.1146 & 8.91 & 75 & 190 & $+83.68$ & 0.21 & RHB \nl
HD~110281  & 6150 & 3367.0801 & 9.39 & 45 & 170 & $+139.90$ & 0.51 & RGB; KR~Vir; jitter \nl
HD~165195  & 6150 & 4011.7779 & 7.34 & 120 & 380 & $+0.50$ & 0.40 & RGB; V2564~Oph; jitter \nl
HD~184266  & 5430 & 4016.7436 & 7.57 & 50 & 250 & $-349.20$ & 0.39 & RHB \nl
HD~187111  & 6150 & 4015.8079 & 7.75 & 40 & 395 & $-186.16$ & 0.15 & RGB \nl
HD~195636  & 5430 & 4016.7877 & 9.57 & 140 & 130 & $-258.34$ & 1.62 & RHB \nl
HD~214925  & 6150 & 4015.7595 & 9.30 & 50 & 215 & $-327.26$ & 0.60 & RGB; jitter \nl
HD~214362  & 5430 & 4016.9199 & 9.10 & 60  & 105 & $-92.48$  & 0.14 & RHB \nl
HD~218732  & 6150 & 4015.8478 & 8.47 & 40 & 270 & $-294.24$ & 0.25 & RGB; LS~Aqr; SLSB; jitter \nl
HD~221170  & 6150 & 4009.8693 & 7.71 & 100 & 260 & $-121.69$ & 0.40 & RGB; NSV~14589 \nl
$\eta$~Ser & 6150 & 3981.7138 & 3.26 & 20 & 300 & $+12.24$ & 0.15 & RGB; check; NSV~10675 \nl
\tablevspace{4pt}
\enddata
\tablenotetext{a}{S/N values are per pixel. A typical resolution element
is 2.3 pixels.}
\end{deluxetable}

\clearpage

\begin{deluxetable}{lrrrrrrrl}
\tablewidth{0pc}
\tablenum{2}
\footnotesize
\tablecaption{Line Broadening Results \label{tab2}}
\tablehead{
\colhead{Star} &
\colhead{$M_{\rm V}$} &
\colhead{$T_{\rm eff}$} &
\colhead{log $g$} &
\colhead{[Fe/H]} &
\colhead{$V_{\rm broad}$} &
\colhead{$V_{\rm rot}$ sin $i$} &
\colhead{$\zeta_{\rm RT}$} &
\colhead{Comments} }
\startdata
HD~3008 & $-1.5$ & 4140 & 1.00 & $-1.43$ & 9.2 & $4.4 \pm 0.8$ & $6.9 \pm 0.6$ & RGB \nl
BD$-18$~271 & $-2.1$ & 4150 & 0.70 & $-1.98$ & 7.3 & $0.0 \pm 1.5$ & $7.2 \pm 1.0$ & RGB \nl
CD$-36$~1052 & $+0.62$ & 5890 & 2.50 & $-2.00$ & 14.4 & $8.8 \pm 0.8$ & $8.9 \pm 0.8$ & RHB \nl
HD~23798 & $-1.8$ & 4310 & 1.00 & $-1.90$ & 5.0 & $0.0 \pm 1.0$ & $6.7 \pm 0.6$ & RGB \nl
HD~25532 & $+0.79$ & 5320 & 2.54 & $-1.33$ & 8.5 & $4.8 \pm 1.0$ & $7.6 \pm 0.7$ & RHB \nl
BD+6~648 & $-1.79$ & 4160 & 0.87 & $-1.82$ & 6.1 & $1.2 \pm 1.5$ & $6.6 \pm 1.0$ & RGB \nl
HD~29574  & $-2.11$ & 3960 & 0.57 & $-2.11$ & 10.2 & $3.7 \pm 1.0$ & $7.4 \pm 0.6$ & RGB \nl
          & &      &      &         &      & $4.2 \pm 0.7$ & $7.3 \pm 0.5$ & \nl
HD~82590  & $+0.7$ & 5960 & 2.70 & $-1.85$ & 13.0 & $7.7 \pm 0.6$ & $11.0 \pm 0.5$ & RHB \nl
BD+22~2411 & $-1.7$ & 4320 & 1.00 & $-1.95$ & 7.3 & $0.0 \pm 2.0$ & $7.5 \pm 0.7$ & RGB \nl
HD~106373 & $+0.57$ & 6160 & 2.70 & $-2.48$ & 13.5 & $10.8 \pm 0.7$ & $6.5 \pm 1.5$ & RHB \nl
HD~110281 & $-2.6$ & 3850 & 0.20 & $-1.75$ & 11.5 & $5.5 \pm 1.0$ & $6.2 \pm 1.0$ & RGB \nl
HD~165195 & $-2.14$ & 4200 & 0.76 & $-2.16$ & 7.6 & $1.8 \pm 0.7$ & $6.4 \pm 0.5$ & RGB \nl
HD~184266 & $+0.7$ & 5490 & 2.60 & $-1.87$ & 11.7 & $5.0 \pm 0.5$ & $11.5 \pm 0.3$ & RHB \nl
HD~187111 & $-1.54$ & 4260 & 1.04 & $-1.65$ & 5.2 & $2.4 \pm 0.5$ & $5.5 \pm 0.7$ & RGB \nl
HD~195636 & $+0.5$ & 5370 & 2.40 & $-2.40$ & 21.5 & $22.2 \pm 1.0$ & $10.0 \pm 1.5$ & RHB \nl
HD~214925 & $-2.5$ & 3890 & 0.30 & $-2.14$ & 9.7 & $4.5 \pm 0.7$ & $8.4 \pm 0.4$ & RGB \nl
HD~214362 & $+0.6$ & 5700 & 2.60 & $-2.20$ & 11.1 & $7.5 \pm 1.0$ & $8.5 \pm 0.7$ & RHB \nl
HD~218732 & $-2.8$ & 3900 & 0.20 & $-2.00$ & 11.1 & $3.1 \pm 0.5$ & $6.4  \pm 0.7$ & RGB \nl
HD~221170 & $-1.7$ & 4410 & 1.10 & $-1.56$ & 7.4 & $1.0 \pm 1.0$ & $6.4 \pm 0.5$ & RGB \nl
$\eta$~Ser & $+1.87$ & 4890 & 3.21 & $-0.42$ & 4.0 & $1.0 \pm 0.8$ & $4.1 \pm 0.5$ & disk RGB \nl
\tablevspace{4pt}
\enddata
\end{deluxetable}

\clearpage

\begin{deluxetable}{lllrrrrrr}
\tablewidth{0pc}
\tablenum{3}
\footnotesize
\tablecaption{Disk Giants \label{tab3}}
\tablehead{
\colhead{Star} & 
\colhead{HR} &
\colhead{Sp Type} &
\colhead{[Fe/H]} &
\colhead{$T_{\rm eff}$} &
\colhead{log $g$} &
\colhead{$M_{\rm V}$} &
\colhead{$V_{\rm rot}$ sin $i$} &
\colhead{$\zeta_{\rm RT}$} }
\startdata
$\xi$ Cyg & 8079 & K4.5 Ib & $-0.45$ & 4090 & 1.42 & $-4.60_{-0.37}^{+0.45}$ & $1.6 \pm 2.0$ & $10.1 \pm 1.0$ \nl
58 Per & 1454 & G8 II & $-0.29$ & 4260 & 2.21 & $-2.22_{-0.38}^{+0.47}$ & $6.3 \pm 1.4$ & $10.2 \pm 0.9$ \nl
$\theta$ Lyr & 7314 & K0 II & $-0.01$ & 4500 & 1.93 & $-2.76_{-0.24}^{+0.27}$ & $3.6 \pm 1.4$ & $5.6 \pm 0.9$ \nl
$\theta$ Her & 6695 & K1 II & $-0.24$ & 4330 & 1.28 & $-2.71_{-0.23}^{+0.26}$ & $3.4 \pm 0.6$ & $7.9 \pm 0.2$ \nl
$\lambda$ Lyr & 7192 & K2.5 II & $-0.02$ & 4220 & 2.21 & $-3.75_{-0.50}^{+0.65}$ & $3.2 \pm 1.0$ & $6.3 \pm 0.5$ \nl
$\pi$ Her & 6418 & K3 II & $-0.18$ & 4100 & 1.68 & $-2.10_{-0.12}^{+0.13}$ & $3.7 \pm 0.1$ & $4.9 \pm 0.6$ \nl
$\beta$ Cyg & 7417 & K3 II & $-0.17$ & 4270 & 1.79 & $-2.27_{-0.14}^{+0.15}$ & $3.0 \pm 0.3$ & $6.1 \pm 0.2$ \nl
$\gamma$ Aql & 7525 & K3 II & $-0.29$ & 4210 & 1.63 & $-3.38_{-0.22}^{+0.24}$ & $3.2 \pm 0.5$ & $7.0 \pm 0.4$ \nl
1 Lac & 8498 & K3 II-III & $-0.12$ & 4350 & 1.75 & $-2.61_{-0.24}^{+0.27}$ & $3.6 \pm 1.4$ & $5.6 \pm 0.9$ \nl
$\alpha$ Hya & 3748 & K3 II-III & $-0.12$ & 4120 & 1.77 & $-1.68_{-0.09}^{+0.09}$ & $0.0 \pm 1.4$ & $6.5 \pm 0.9$ \nl
$\beta$ Sge & 7488 & G8 IIIa & $-0.03$ & 4850 & 2.79 & $-1.39_{-0.20}^{+0.22} $ & $9.1 \pm 0.7$ & $8.2 \pm 0.7$ \nl
$\lambda$ Peg & 8667 & G8 IIIa & $-0.10$ & 4800 & 3.20 & $-1.46_{-0.18}^{+0.19}$ & $7.8 \pm 0.4$ & $7.8 \pm 0.4$ \nl
$\beta$ Her & 6148 & G8 III & $-0.27$ & 4920 & 2.62 & $-0.49_{-0.10}^{+0.10}$ & $3.4 \pm 1.0$ & $6.8 \pm 1.0$ \nl
$\eta$ Dra & 6132 & G8 III & $-0.21$ & 4940 & 3.10 & $+0.59_{-0.03}^{+0.03}$ & $2.2 \pm 1.0$ & $5.5 \pm 1.0$ \nl
$\alpha^{2}$ Cap & 7754 & G8 IIIb & $-0.18$ & 5000 & 3.05 & $+0.98_{-0.06}^{+0.07}$ & $3.2 \pm 0.4$ & $4.6 \pm 0.3$ \nl
$\alpha$ Cas & 168 & K0 III & $-0.09$ & 4610 & 2.71 & $-1.98_{-0.09}^{+0.09}$ & $4.9 \pm 0.4$ & $6.2 \pm 0.3$ \nl
$\delta$ Tau & 1373 & K0 III & 0.00 & 4940 & 2.85 & $+0.40_{-0.09}^{+0.10}$ & $2.5 \pm 1.0$ & $6.2 \pm 1.0$ \nl
$\gamma$ Tau & 1346 & K0 III & $-0.02$ & 4930 & 2.90 & $+0.28_{-0.12}^{+0.12} $ & $2.4 \pm 1.0$ & $5.9 \pm 1.0$ \nl
$\beta$ Cet & 188 & K0 III & $-0.09$ & 4820 & 2.87 & $-0.30_{-0.05}^{+0.05}$ & $3.0 \pm 1.0$ & $5.9 \pm 1.0$ \nl
$\theta^{1}$ Tau & 1411 & K0 III & $+0.04$ & 4960 & 3.17 & $+0.43_{-0.09}^{+0.09}$ & $3.4 \pm 1.0$ & $4.9 \pm 1.0$ \nl
$\beta$ Gem & 2990 & K0 III & $-0.04$ & 4850 & 2.75 & $+1.08_{-0.02}^{+0.02}$ & $2.5 \pm 1.0$ & $4.2 \pm 1.0$ \nl
$\epsilon$ Cyg & 7949 & K0 III & $-0.27$ & 4730 & 2.89 & $+0.78_{-0.03}^{+0.03}$ & $3.0 \pm 1.0$ & $4.2 \pm 1.0$ \nl
$\alpha$ UMa & 4301 & K0 III & $-0.20$ & 4660 & 2.46 & $-1.10_{-0.04}^{+0.04}$ & $2.6 \pm 1.0$ & $5.2 \pm 1.0$ \nl
$\gamma^{1}$ Leo & 4057 & K2 III & $-0.49$ & 4470 & 2.35 & $-0.32_{-0.07}^{+0.07}$ & $2.6 \pm 1.0$ & $5.2 \pm 1.0$ \nl
$\alpha$ Ari & 617 & K2 III & $-0.25$ & 4480 & 2.57 & $+0.47_{-0.04}^{+0.04}$ & $3.1 \pm 1.0$ & $3.9 \pm 1.0$ \nl
$\beta$ Oph & 6603 & K2 III & $+0.02$ & 4550 & 2.63 & $+0.77_{-0.04}^{+0.04}$ & $1.6 \pm 1.0$ & $4.0 \pm 1.0$ \nl
$\alpha$ Ser & 5854 & K2 III & $+0.03$ & 4530 & 2.76 & $+0.88_{-0.03}^{+0.03}$ & $0.0 \pm 1.0$ & $4.8 \pm 1.0$ \nl
$\alpha$ Boo & 5340 & K2 III & $-0.60$ & 4200 & 2.19 & $-0.30_{-0.02}^{+0.02}$ & $2.4 \pm 1.0$ & $5.2 \pm 1.0$ \nl
$\eta$ Ser & 6869 & K2 III & $-0.42$ & 4890 & 3.21 & $+1.87_{-0.03}^{+0.03}$ & $2.0 \pm 0.5$ & $4.0 \pm 0.5$ \nl 
$\epsilon$ Sco & 6241 & K2.5 III & $-0.17$ & 4560 & 2.49 & $+0.78_{-0.04}^{+0.04}$ & $2.6 \pm 0.5$ & $4.3 \pm 0.5$ \nl
$\epsilon$ Crv & 4630 &  K2.5 III & $+0.13$ & 4320 & 2.16 & $-1.82_{-0.14}^{+0.15}$ & $2.6 \pm 0.5$ & $5.8 \pm 0.5$ \nl
26 Aql & 7333 & G8 III-IV & $-0.21$ & 4900 & 3.31 & $+1.63_{-0.08}^{+0.08}$ & $2.8 \pm 0.5$ & $5.0 \pm 0.5$ \nl
\tablevspace{4pt}
\enddata
\end{deluxetable}

\clearpage

\begin{deluxetable}{lrrrrrrrrr}
\tablewidth{0pc}
\tablenum{4}
\footnotesize
\tablecaption{The Other 13 Red Horizontal Branch Stars \label{tab4}}
\tablehead{
\colhead{Star} &
\colhead{$\alpha$ (J2000)} &
\colhead{$\delta$ (J2000)} &
\colhead{[Fe/H]} &
\colhead{$M_{V}$} &
\colhead{$T_{eff}$} &
\colhead{log~$g$} &
\colhead{$R$/$R_{\odot}$} &
\colhead{$V_{\rm broad}$} &
\colhead{$V_{\rm rot}$\tablenotemark{a}}}
\startdata
HD 20 & 00:05:15.3 & $-27$:16:18 & $-1.66$ & 0.7 & 5350 & 2.5 & 8.2 
 & 5.9 & 0.8 \nl
CD$-23$ 72 & 00:16:16.5 & $-22$:34:40 & $-1.12$ & 0.8 & 5270 & 2.5 & 8.0 & 8.9 & 3.1 \nl
HD 3179 & 00:34:50.6 & $-21$:52:56 & $-0.92$ & 0.9 & 5280 & 2.6 & 7.8 &
 5.2 & 0.4 \nl
BD$+44$ 493 & 02:26:49.7 & +44:57:46 & $-2.71$ & 0.8 & 5510 & 2.6 & 7.4 & 3.9 & 0.0 \nl
HD 108317 (binary) & 12:24:04.4 & +05:34:46 & $-2.48$ & 0.5 & 5230 & 2.4 & 9.6 &
 5.1 & 0.4 \nl
HD 110885 & 12:45:19.2 & +01:03:20 & $-1.59$ & 0.7 & 5330 & 2.5 & 8.2 &
 8.2 & 2.5 \nl
HD 119516 & 13:43:26.7 & +15:34:29 & $-2.49$ & 0.6 & 5440 & 2.5 & 8.6 &
 9.1 & 3.3 \nl
BD$+9$ 2860 & 14:13:19.7 & +08:36:40 & $-1.67$ & 0.7 & 5240 & 2.5 & 8.6  & 
 3.9 & 0.0 \nl
BD$+11$ 2998 & 16:30:16.7 & +10:59:51 & $-1.46$ & 0.8 & 5360 & 2.5 & 8.0 & 
 6.8 & 1.4 \nl
BD$+9$ 3223 & 16:33:35.6 & +09:06:17 & $-2.41$ & 0.6 & 5310 & 2.4 & 8.9 & 
 4.8 & 0.2 \nl
BD+17 3248   & 17:28:14.4 & $+17$:30:35  & $-2.07$ & 0.65    & 5240 & 2.44 &  9.0 & 4.3 & 0.0 \nl
BD+25 3410   & 18:02:03.2 & $+25$:00:41  & $-1.37$ & 0.79    & 5740 & 2.69 &  6.7 & 9.7 & 3.9 \nl
BD$-3$ 5215 & 21:28:01.3 & $-03$:07:40 & $-1.64$ & 0.7 & 5420 & 2.6 & 7.9 & 
 7.3 & 1.8 \nl
\tablevspace{4pt}
\tablenotetext{a}{These values are derived using Equation~\ref{eq:vbroadcorr}.}
\enddata
\end{deluxetable}

\clearpage

\begin{deluxetable}{lrrrrrrrrrrr}
\tablewidth{0pc}
\tablenum{5}
\footnotesize
\tablecaption{CfA Data Binned by $M_{\rm V}$ \label{tab5}}
\tablehead{
\colhead{$<M_{\rm V}>$} &
\colhead{N} &
\colhead{$\sigma$} &
\colhead{$\sigma_{\mu}$} &
\colhead{$<R>$} &
\colhead{$\sigma$} &
\colhead{$\sigma_{\mu}$} &
\colhead{$<V_{\rm broad}>$} &
\colhead{$\sigma$} &
\colhead{$\sigma_{\mu}$} &
\multicolumn{2}{r}{$<V_{\rm rot}$ sin $i>$} \\
\colhead{} & \colhead{} & \colhead{} & \colhead{} & \colhead{($R_{\odot}$)} & \colhead{} &
\colhead{} & \colhead{} & \colhead{} & \colhead{} & \colhead{Eq. 1} & \colhead{Eq. 2} }
\startdata
 $-2.01$ & 20 & 0.52 & 0.12 & 64.3 & 21.7 & 4.8 & 7.7 & 2.3 & 0.5 & 4.4 & 2.1 \nl
 $-1.28$ & 20 & 0.37 & 0.08 & 34.9 & 5.8 & 1.3 & 5.2 & 1.7 & 0.4 & 0.0 & 0.4 \nl
 $-0.67$ & 20 & 0.38 & 0.09 & 22.2 & 3.4 & 0.8 & 5.5 & 1.7 & 0.4 & 0.0 & 0.6 \nl
 $+0.07$ & 20 & 0.32 & 0.07 & 14.7 & 2.3 & 0.5 & 4.2 & 1.4 & 0.3 & 0.0 & 0.0 \nl
 $+0.83$ & 20 & 0.50 & 0.11 & 9.4 & 1.0 & 0.2 & 3.8 & 2.2 & 0.5 & 0.0 & 0.0 \nl
 $+1.89$ & 13 & 0.68 & 0.19 & 5.4 & 1.0 & 0.3 & 2.9 & 2.5 & 0.7 & 0.0 & 0.0 \nl
\tablevspace{4pt}
\enddata
\end{deluxetable}

\clearpage

\begin{deluxetable}{lrrrrrrrr}
\tablewidth{0pc}
\tablenum{6}
\footnotesize
\tablecaption{Comparison of Derived Rotational Velocities with Mg~II and
Ca~II Emission line Asymmetries \label{tab6}}
\tablehead{
\colhead{Star} &
\colhead{$M_{\rm V}$} &
\colhead{$V_{\rm rad}$} &
\colhead{$V_{\rm rot}$ sin $i$} &
\colhead{$V_{\rm broad}$(CfA)} &
\colhead{$V_{\rm rot}$(Eq.\ref{eq:vbroad})} &
\colhead{$V_{\rm rot}$(Eq.\ref{eq:vbroadcorr})} &
\colhead{Mg~II} &
\colhead{Ca~II} }
\startdata
HD 2796  & $-0.81$ & $ -61.0$ & \nodata & 7.0 & 4.1 & 1.6 & \nodata & V$>$R \nl
HD 6755\tablenotemark{a}  &  1.50 & $-319.2$ & \nodata &  3.3 & 0.0 & 0.0 & V$>$R & V$>$R  \nl
HD 6833  & $-0.40$ & $-245.0$ & \nodata & 7.4 & 5.0 & 1.9 & V$<$R & V=R \nl
HD 8724  & $-1.11$ & $-113.2$ & \nodata & 6.0 & 1.4 & 0.9 & \nodata &  V$>$R \nl
HD 21022 & $-1.17$ & 122.3 & \nodata & 5.0 & 0.0 & 0.3 & \nodata & V$>$R \nl
HD 23798 & $-1.80$ & 89.4 & 0.0 & 5.0 & 0.0 & 0.3 & \nodata & V=R \nl
HD 26297\tablenotemark{b} & $-1.48$ & 14.8 & \nodata & 0.5 & 0.0 & 0.0 & \nodata & V$>$R  \nl
HD 29574\tablenotemark{c} & $-2.11$ & 19.8 & 4.0 & 10.1 & 7.8 & 4.3 & \nodata & V$<$R \nl
HD 36702 & $-1.90$ & 122.7 & \nodata & 6.5 & 1.5 & 1.2 & \nodata & V$>$R \nl
HD 63791 &  0.30 & $-108.4$ & \nodata & 3.7 & 0.0 & 0.0 & \nodata & V$>$R \nl
HD 83212 & $-0.90$ & 109.1 & \nodata & 7.3 & 4.5 & 1.8 & \nodata & V$<$R \nl
HD 88609 & $-1.42$ & $-38.5$ & \nodata & 4.0 & 0.0 & 0.0 & \nodata & V$>$R \nl
HD 110281 & $-2.60$ & 141.9 & 5.5 & 11.5 & 9.3 & 5.8 & \nodata & V=R \nl
HD 118055  & $-1.50$ & $-100.7$ & \nodata & 7.4 & 4.2 & 1.9 & \nodata & V$<$R \nl
HD 122956\tablenotemark{d} & $-0.70$ & 166.0 & \nodata & 6.3 & 2.9 & 1.1 & V$<$R & V$>$R \nl
HD 126587  & $-0.60$ & 149.0 & \nodata & 4.3 & 0.0 & 0.0 & V$>$R & \nodata \nl
BD+1 2916  & $-1.76$ & $-12.1$ & \nodata & 8.2 & 5.3 & 2.5 & \nodata & V$>$R \nl
HD 166161  & 0.79 & 68.3 & \nodata & 4.0 & 0.0 & 0.0 & \nodata & V$>$R \nl
HD 175305  & 1.80 & $-181.0$ & \nodata & 7.1 & 5.8 & 1.7 & V$>$R & V=R \nl
HD 184711\tablenotemark{e}  & $-2.35$ & 102.2 & \nodata & 7.7 & 4.0 & 2.1 & \nodata & V$\leq$R \nl
HDE 232078 & $-2.15$ & $-387.2$ & \nodata & 10.9 & 8.8 & 5.2 & \nodata & V=R \nl
HD 187111\tablenotemark{f} & $-1.54$ & $-186.5$ & 2.4 & 5.2 & 0.0 & 0.4 & \nodata & V$\leq$R \nl
HD 204543\tablenotemark{g}  & $-1.09$ & $-98.4 $ & \nodata & 5.0 & 0.0 & 0.3 & \nodata &  V$\geq$R \nl
HD 216143  & $-1.4$ & $-116.0$ & \nodata & 5.5 & 0.0 & 0.6 & V$<$R &  V$>$R \nl
HD 218732\tablenotemark{h} & $-2.80$ & $-312.2$ & 3.1 & 11.1 & 8.7 & 5.4 & \nodata & V$<$R \nl
HD 221170  & $-1.70$ & $-119.0$ & 1.0 & 7.4 & 4.0 & 1.9 & V$<$R & V$>$R \nl
HD 222434\tablenotemark{i}  & $-1.20$ & 8.8 & \nodata & 5.0 & 0.0 & 0.3 & \nodata & V=R \nl
\tablevspace{4pt}
\enddata
\tablenotetext{a}{We employ the $\gamma$ velocity for this binary system.}
\tablenotetext{b}{Dupree \& Smith 1995 cite
$V_{\rm rad}$ = +135 \kms, which we believe to be incorrect.}
\tablenotetext{c}{Smith et al.\ 1992
and Dupree \& Smith 1995 both found V$>$R.}
\tablenotetext{d}{Smith et al.\ 1992 and Dupree \&
Smith 1995 both found V$>$R.} 
\tablenotetext{e}{Smith et al.\ 1992 found V$<$R while Dupree \&
Smith 1995 found it to be uncertain. We adopt the former value.} 
\tablenotetext{f}{Smith et al.\ 1992
found V$<$R while Dupree \& Smith  1995 found V=R.}
\tablenotetext{g}{Smith et al.\ 1992
found V$>$R and Dupree \& Smith 1995 found V$\approx$R. Both suggested
$M_{\rm V}$ = $-0.3$, but we recommend $-1.09$.}
\tablenotetext{h}{We employ the $\gamma$ velocity
for this binary system.} 
\tablenotetext{i}{We employ the $\gamma$ velocity for this binary system.}
\end{deluxetable}

\clearpage

\end{document}